\documentclass[preprint,onecolumn,superscriptaddress]{revtex4}

\usepackage{amssymb}
\usepackage{enumerate}
\usepackage{amssymb}
\usepackage{amsmath}
\usepackage{feynmf}
\usepackage{slashed}
\usepackage{natbib}
\usepackage{graphicx}
\usepackage[pdftex,bookmarks,linktocpage,pdfpagelabels,plainpages=false,hyperfigures,linkcolor=blue,citecolor=blue]{hyperref} 
\hypersetup{colorlinks=true}

\usepackage{appendix}

\usepackage{mathrsfs,amssymb}  
\usepackage{cancel}
\usepackage[normalem]{ulem}

\newcommand\be{\begin{equation}}
\newcommand\ee{\end{equation}}

\newcommand\bea{\begin{eqnarray}}
\newcommand\eea{\end{eqnarray}}

\begin{document}
\begin{flushright}
MI-TH-2026
\end{flushright}
\bibliographystyle{apsrev4-1}

\title{Present and future status of light dark matter models from cosmic-ray electron upscattering}

\author{James B.~Dent} 
\affiliation{Department of Physics, Sam Houston State University, Huntsville, TX 77341, USA}

\author{Bhaskar Dutta}
\affiliation{Mitchell Institute for Fundamental Physics and Astronomy,
   Department of Physics and Astronomy, Texas A\&M University, College Station, TX 77845, USA}

\author{Jayden L.~Newstead}
\affiliation{ARC Centre of Excellence for Dark Matter Particle Physics, School of Physics, The University of Melbourne, Victoria 3010, Australia}

\author{Ian M. Shoemaker}
\affiliation{Center for Neutrino Physics, Department of Physics, Virginia Tech University, Blacksburg, VA 24601, USA}

\author{Natalia Tapia Arellano}
\affiliation{Center for Neutrino Physics, Department of Physics, Virginia Tech University, Blacksburg, VA 24601, USA}

\begin{abstract}

Non-relativistic Dark Matter (DM) can be accelerated by scattering on high-energy cosmic-ray (CR) electrons. This process leads to a sub-population of relativistic or semi-relativistic DM which extends the experimental reach for direct detection in the sub-GeV mass regime. In this paper we examine the current and future potential of this mechanism for constraining models of light dark matter. In particular, we find that  Super-Kamiokande and XENON1T data can already provide leading constraints on the flux of dark matter that has been accelerated to high energies from cosmic ray electrons. We also examine future projected sensitivities for DUNE and Hyper-K, and contrary to previous findings, conclude that DUNE will be able supersede Super-K bounds on cosmic-ray upscattered DM for a variety of DM models. 


\end{abstract}

\maketitle

\section{Introduction}
The nature of the non-luminous dark matter (DM) is one of the greatest mysteries in physics. At present, everything that is known about DM has been derived on the basis of its gravitational interactions with ordinary matter. The race is on to detect any non-gravitational interactions of DM, which would be a significant step forward in our understanding of the most abundant type of matter in the Universe. 

An extremely active and promising direction involves ``direct detection'' searches which aim to observe small energy depositions within underground detectors~\cite{Goodman:1984dc} via scattering on nuclei or electrons. 
These experiments lose sensitivity to low-mass DM ($m_\chi \lesssim\mathcal{O}$ GeV) given that the $\mathcal{O}(100)~{\rm km}/{\rm s}$ DM velocities in the Milky Way limit the available energy to be deposited in a detector. 

A striking caveat to this conclusion comes from the presence of high-energy cosmic rays. 
If DM possess a nonzero cross section with ordinary matter, high-energy cosmic rays can scatter the relatively slow-moving DM and thereby ``boost'' a fraction of the galactic DM above the critical energy necessary to deposit energy above the detectable threshold at a direct detection experiment. This extends the reach of standard direct detection experiments orders of magnitude below the typical $\mathcal{O}$(GeV) mass reach. The enhanced sensitivity at lower masses does come at a cost, however, as the flux from the upscattered subset is substantially less than the general galactic DM population, which necessitates larger cross sections for observability.  

The possibility of examining sub-GeV mass dark matter through CR upscattering has recently received significant attention~\cite{Ema:2018bih,Bringmann:2018cvk,Dent:2019krz,Cappiello:2019qsw}, and can also produce observable effects on the CR spectrum itself via DM induced energy losses~\cite{Cappiello:2018hsu}. In addition to the possibility of extending the reach of conventional direct detection experiments, these initial studies demonstrated that an upscattered DM population could also be probed with a variety of neutrino experiments such as DAYA BAY, PROSPECT, Borexino, and Super-Kamiokande, with future possibilities for JUNO, DUNE, and Hyper-K. Neutrino experiments are able to provide stringent bounds on cosmic-ray upscattered DM due to the large exposures at these experiments and the increase in incident high-energy DM flux from the CR scattering process. The parameter space probed by these experiments is smaller in mass and considerably larger in cross section than the peak sensitivities of direct detection experiments which typically probe the 10~GeV$-$1~TeV mass range at cross sections of $\mathcal{O}(10^{-46}-10^{-44}$~cm$^2$).

Other recent methods for improving the bounds on light DM provide complementary probes using e.g. the Migdal effect~\cite{Ibe:2017yqa,Dolan:2017xbu,Akerib:2018hck,Armengaud:2019kfj,Bell:2019egg,Liu:2019kzq}, Bremsstrahlung~\cite{Kouvaris:2016afs}, or CMB distortions~\cite{Gluscevic:2017ywp,Boddy:2018kfv,Boddy:2018wzy}. Cosmic-ray upscattering of DM has also received renewed interested in the context of the recent XENON1T excess~\cite{Su:2020zny,Cao:2020bwd,Bloch:2020uzh,Zhang:2020htl}. 

Previous work on CR upscattered DM through electron interactions has focused on energy independent cross-sections \cite{Ema:2018bih,Cappiello:2019qsw}. However, the importance of energy-dependent scattering has been recently highlighted in the context of nuclear recoil experiments~\cite{Dent:2019krz}, where it was found that the resulting cross section bounds can be orders of magnitude different than those derived under the assumption of a constant cross section. Therefore, in this paper we revisit the direct detection of CR upscattered DM via electron recoil in specific models of DM interactions and thereby extend the work in~\cite{Dent:2019krz}. We examine the reach of Super-K and XENON1T for a variety of interaction types including anapole, magnetic dipole, pseudoscalar mediated, and leptophilic scalar mediated interactions. We find that Super-K produces leading constraints on a wide variety of DM models at low masses. Moreover, the upcoming experiments DUNE and Hyper-K will significantly improve these bounds. 

The remainder of this work is organized as follows. In Sec.~\ref{sec:calculationframework} we discuss how to obtain the DM flux at and event rate within a terrestrial detector from an upscattered DM population. In Sec.~\ref{sec:models} the specific models of dark matter-electron interactions to be investigated are described. Sec.~\ref{sec:sensitivities} provides the results in the cross-section vs. dark matter mass plane from Super-K and XENON1T, along with projections from DUNE and Hyper-K, with a summary following in Sec.\ref{sec:conclusions}.



\begin{figure}[t!]
\includegraphics[angle=0,width=.50\textwidth]{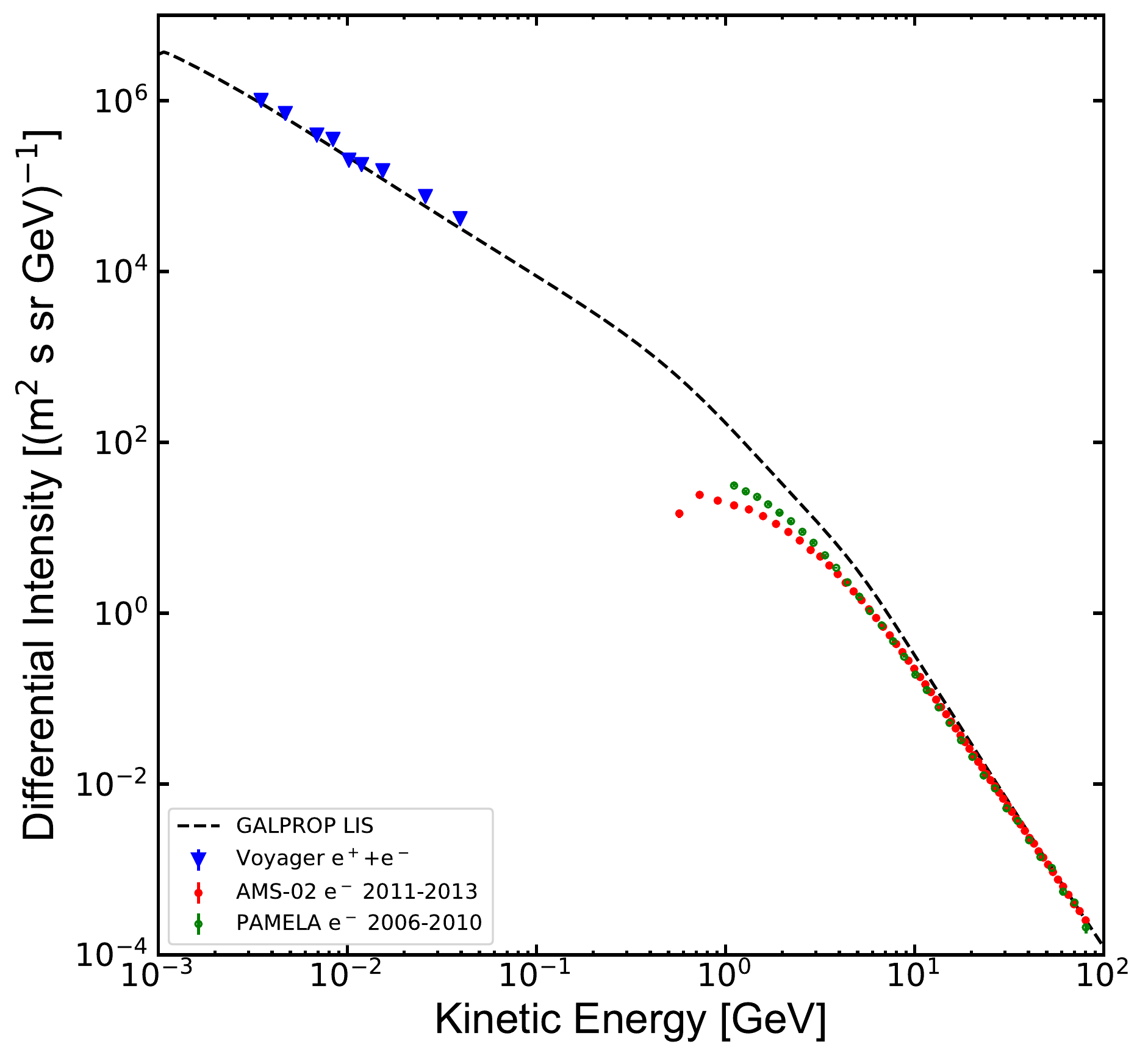}
\caption{Here we reproduce the fit to the electron CR flux from~\cite{Boschini:2018zdv}.}
\label{fig:CRflux}
\end{figure}

\section{Calculational Framework}
\label{sec:calculationframework}

Here we describe the essential set of assumptions in order to compute event rates of CR upscattered DM at an experiment.  As we will see, although there are important astrophysical uncertainties, the experimental sensitivity we obtain will not be strongly impacted by them. This is due to the fact that the rate of events in an experiment scales as two powers of the dark matter-electron cross section, $\propto \sigma_{\chi e}^{2}$, one power each from the upscattering and detection processes.

The double-differential collision rate of CR electrons with dark matter within an infinitesimal volume is
\bea
\frac{d^2\Gamma_{\rm{CR}_e\rightarrow\chi}}{dT_edT_\chi} = \frac{\rho_\chi}{m_\chi} \frac{d\sigma_{\chi e}}{dT_\chi} \frac{d\Phi^{\rm{LIS}}_e}{dT_e} dV
\eea
the scattered dark matter flux is then obtained by integrating this over the relevant volume and CR energies
\bea
\frac{d\Phi_\chi}{dT_\chi} &=& \int_V\!dV\int_{T_e^{\rm{min}}}\!\!\!\!dT_e \,\,\frac{d^2\Gamma_{\rm{CR}_e\rightarrow\chi}}{dT_idT_\chi}\\
&=& D_{\rm eff}\,\frac{\rho_\chi}{m_\chi}\,  \int_{T_e^{\rm{min}}}\!\!\!\!dT_e\,\frac{d\sigma_{\chi e}}{dT_\chi}\,\frac{d\Phi^{\rm{LIS}}_e}{dT_i} 
\eea
where $D_{{\rm eff}}$ is an effective diffusion zone parameter, $\rho_{\chi} = 0.3~{\rm GeV}{\rm cm}^{-3}$ is the local DM density, and $d\Phi^{\rm{LIS}}_i/dT_i$ is the local interstellar (LIS) flux of electrons. We use the fluxes of CRs in Ref.~\cite{Boschini:2018zdv,Bisschoff:2019lne}, which are obtained from a fit to Voyager 1~\cite{Cummings_2016}, AMS-02~\cite{Aguilar:2014fea}, and PAMELA data~\cite{Adriani:2011xv}.

In this calculation the diffusion zone sets the distance from Earth over which we include contributions to the high-energy DM flux. The parameter $D_{{\rm eff}}$ takes into account the variation of the dark matter density throughout the diffusion zone. While there is uncertainty in the precise value of this parameter, we follow Ref.~\cite{Bringmann:2018cvk} and conservatively consider a diffusion zone of 1 kpc, which corresponds to $D_{{\rm eff}} = 0.997$ kpc. 

Finally, we obtain the differential event rate (per unit detector mass) from the incoming relativistic DM flux as measured at a terrestrial detector
\be 
\frac{dR}{dE_{T}} =\frac{1}{m_{N}} \int_{T_{\chi}^{{\rm min}}}^{\infty} dT_{\chi}~\frac{d\Phi_\chi}{dT_\chi}~ \frac{d \sigma_{\chi-e}}{dE_{T}}
\label{eq:rate}
\ee
where $E_{T}$ is the recoil energy of the target electron.

\begin{figure}[t!]
\includegraphics[angle=0,width=.50\textwidth]{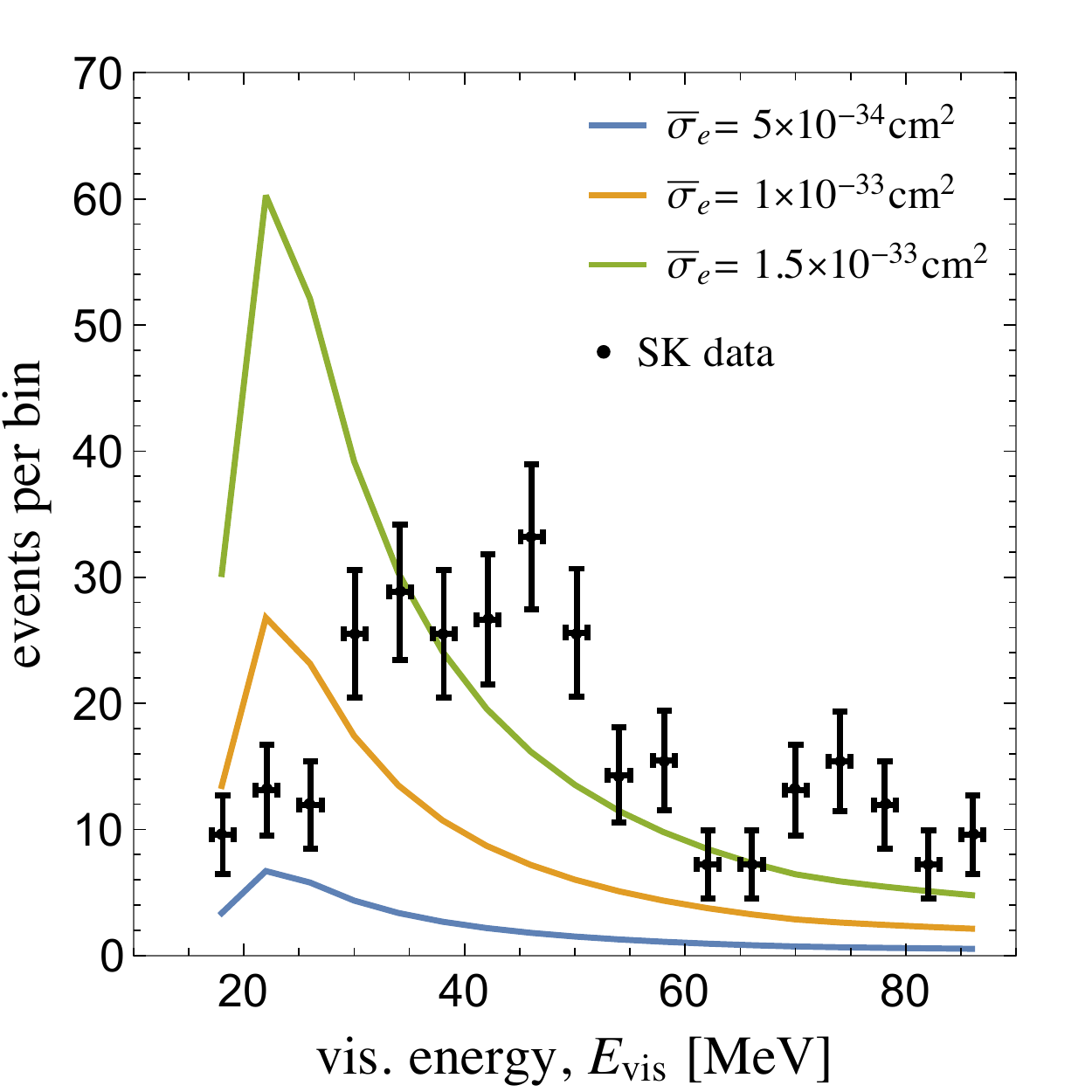}
\caption{Here compare the predicted event rates for a variety of cross sections in the magnetic dipole interaction case and compare them to existing Super-K data~\cite{Bays:2011si}. In this example, the DM mass has been fixed to $m_{X}=10^{-5}$ GeV. }
\label{fig:SKcompare}
\end{figure}

\section{Models of DM-electron Interaction}
\label{sec:models}

In this section we will deduce the non-relativistic reference cross-sections for a variety of interactions. In order to go to the non-relativistic limit for incoming dark matter scattering electrons on earth, for the four-momentum transfer we will use $q^2 = 2m_eE_T \simeq -|\vec{q}|^2$ which we replace with the reference momentum $|\vec{q}|^2 = \alpha^2m_e^2$, as is standard in the field \cite{Essig:2011nj,Essig:2015cda,Trickle:2019ovy}, to find the matrix element $|\mathcal{M}_{\rm{free}}|^2$. We will drop kinetic terms in comparison with mass terms, and use the notation $g_\chi$ and $g_e$ to denote couplings of the mediator to dark matter and electrons, respectively. In each interaction scenario we reduce to the matrix element squared $|\mathcal{M}_{\rm{free}}|^2$ and use the definition \cite{Essig:2011nj,Essig:2015cda,Trickle:2019ovy}
\bea
\bar{\sigma}_e = \frac{\mu_{\chi e}^2}{16\pi m_\chi^2m_e^2}|\mathcal{M}_{\rm{free}}|^2,
\eea
where $\mu_{\chi e}$ is DM-electron reduced mass. {Note that in order to place conservative bounds throughout we do not include atomic coherence which enhances the scattering rate at low energies~\cite{Baxter:2019pnz,Liu:2020pat} 

We begin our examination of DM models by looking at a representative set of spin-dependent electron interactions as found in Table I of \cite{Trickle:2019ovy}. This is a useful set of interactions to explore since it is projected that such interactions will be strongly constrained via magnon excitation~\cite{Trickle:2019ovy} in future detectors. We then investigate a leptophilic scalar dark matter model with a scalar mediator.

\subsection{Pseudo-mediated DM}

The pseudo-mediated DM interaction Lagrangian contains the interaction terms
\bea
\mathcal{L} \supset g_\chi\bar{\chi}\chi\phi + g_e\bar{e}i\gamma^5e\phi
\eea
which leads to the relation
\bea
\frac{d\sigma}{dE_T} = \frac{E_T\bar{\sigma}_e(2m_\chi^2+m_eE_T)(\alpha^2m_e^2+m_\phi^2)^2}{2\alpha^2\mu_{\chi e}^2(2m_\chi T_\chi + T_\chi^2)(2m_eE_T + m_\phi^2)^2},
\eea
for the cross section to produce a recoil energy $E_{T}$ in the target from high-eneregy DM with kinetic energy $T_{\chi}$. 

For the upscattering cross-section we get
\bea
\frac{d\sigma}{dT_\chi} = \frac{T_\chi\bar{\sigma}_em_\chi(2m_\chi+T_\chi)(\alpha^2m_e^2+m_\phi^2)^2}{2\alpha^2\mu_{\chi e}^2(2m_e T_i+T_i^2)(2m_xT_\chi + m_\phi^2)^2}
\eea
This result, and those that follow for other interactions, demonstrates the energy and mediator mass dependence of such processes.

\subsection{Magnetic Dipole Moment DM}

In this case interaction Lagrangian is
\bea
\mathcal{L} \supset \frac{g_\chi}{\Lambda_\chi}\bar{\chi}\sigma^{\mu\nu}\chi V_{\mu\nu} + g_e\bar{e}\gamma^\mu eV_\mu
\eea
This can be applied to the differential cross section for upscattering
\bea
\left(\frac{d\sigma}{dT_\chi}\right)_{MDM,\rm{CR}} =&& \frac{\bar{\sigma}_e(m_\chi+m_e)^2(\alpha^2m_e^2+m_v^2)^2}{(6m_\chi^2 + m_e^2)(\alpha^4m_e^4)}
\\\nonumber
&&\times\frac{2m_\chi^2T_\chi(T_\chi(m_\chi-2(m_e+T_i))+2T_i(2m_e+T_i))}{(2m_\chi T_\chi + m_v^2)^2(2m_e T_i + T_i^2)}
\eea
or scattering on earth
\bea
\left(\frac{d\sigma}{dE_{\rm T}}\right)_{MDM,\rm{CR}} =&& \frac{\bar{\sigma}_e(m_\chi+m_e)^2(\alpha^2m_e^2+m_v^2)^2}{(6m_\chi^2 + m_e^2)(\alpha^4m_e^4)}
\\\nonumber
&&\times\frac{2m_eE_T(m_\chi^2E_T + 2m_e T_\chi(2m_\chi + T_\chi) - 2m_eE_T(m_\chi+T_\chi))}{(2m_eE_T + m_v^2)^2(2m_\chi T_\chi + T_\chi^2)}
\eea

\subsection{Anapole Moment DM}

For the anapole moment we will use the interaction
\bea
\mathcal{L} = \frac{g}{\Lambda_2}\bar{\chi}\gamma^\mu\gamma^5\chi\partial^\nu F_{\mu\nu}
\eea

This can be applied to the differential cross-section for upscattering
\bea
\left(\frac{d\sigma}{dT_\chi}\right)_{AnDM,\rm{CR}} =&& \frac{\bar{\sigma}_e(\alpha^2m_e^2+m_v^2)^2}{6\mu_{\chi e}^2(\alpha^4m_e^4)}
\\\nonumber
&&\times \frac{\left(2m_\chi T_i(2m_e+T_i) -T_\chi(m_e^2+2m_\chi(m_e+T_i)-m_\chi^2) +m_\chi T_\chi^2\right)}{(2m_e T_i + T_i^2)}
\eea
or scattering on earth
\bea
\left(\frac{d\sigma}{dE_{\rm T}}\right)_{AnDM,\rm{CR}} =&& \frac{\bar{\sigma}_e(\alpha^2m_e^2+m_v^2)^2}{6\mu_{\chi e}^2(\alpha^4m_e^4)}
\\\nonumber
&&\times \frac{\left(2m_eT_\chi(2m_\chi+T_\chi)-E_T(m_e^2+2m_e(m_\chi+T_\chi)-m_\chi^2)+m_eE_T^2\right)}{(2m_\chi T_\chi + T_\chi^2)}
\eea

\subsection{Leptophilic Scalar Model}


Finally, we examine a model with a leptophilic scalar DM coupled to a scalar mediator, whose constraints are discussed in \cite{Knapen:2017xzo}. The interaction terms are
\bea
\mathcal{L}_{\rm int} \supset -\frac{1}{2}y_\chi m_\chi\phi\chi^2 - y_e\phi\bar{e}e
\eea

Let us first check the non-relativistic version where an incoming DM particle elastically scatters off an electron. After averaging over the initial electron spin, this provides the matrix element
\bea
|\mathcal{M}|^2 = \frac{y_\chi^2m_\chi^2y_e^2}{(t-m_\phi)^2}4m_e^2
\eea
Evaluating this at $t = q^2 \simeq -|\vec{q}|^2 = -(\alpha m_e)^2$ and using the reference cross section definition
\bea
\bar{\sigma}_e = \frac{\mu_{\chi e}^2}{16\pi m_\chi^2m_e^2}|\mathcal{M}|^2\bigg|_{\vec{q} = \alpha m_e}
\eea
we find
\bea
\label{eq:refscalarDM}
\bar{\sigma}_e = \frac{y_\chi^2y_e^2}{4\pi}\frac{\mu_{\chi e}^2}{\left(m_\phi^2 + \alpha^2m_e^2\right)^2}
\eea
This form is precisely that of Eq.~(43) in \cite{Knapen:2017xzo}.

Now let us evaluate this interaction for cosmic-ray electrons scattering on DM in the DM's rest frame. The differential cross-section for the upscattered DM is
\bea
\frac{d\sigma}{dT_{\chi}} = \frac{|\mathcal{M}|^2}{32\pi m_\chi|\vec{p}_1|^2} = \frac{y_\chi^2y_e^2m_\chi m_e^2}{8\pi(t-m_\phi^2)^2|\vec{p}_1|^2}= \frac{y_\chi^2y_e^2m_\chi m_e^2}{8\pi(2m_\chi T_\chi+m_\phi^2)^2(T_i^2 + 2m_eT_i)}
\eea
The differential cross section for the electron target on Earth is
\bea
\frac{d\sigma}{dE_{\rm T}} = \frac{|\mathcal{M}|^2}{32\pi m_e|\vec{p}_1|^2} = \frac{y_\chi^2y_e^2m_\chi^2m_e}{8\pi(t-m_\phi^2)^2|\vec{p}_1|^2}= \frac{y_\chi^2y_e^2m_\chi^2m_e}{8\pi(2m_eE_{\rm T}+m_\phi^2)^2(T_\chi^2 + 2m_\chi T_\chi)}
\eea
We could solve for either the coupling combination or the mediator mass squared in terms of the reference cross section using Eq.(\ref{eq:refscalarDM}).

This can be applied to the differential cross section for upscattering
\be
\left(\frac{d\sigma}{dT_\chi}\right)_{scalar,\rm{CR}} = \frac{\bar{\sigma}_e(\alpha^2m_e^2+m_{\phi}^2)^2 m_{\chi}m_{e}^{2}}{2 \mu_{\chi e}^{2}\left(2m_{\chi}T_{\chi}+m_{\phi}^{2}\right)^{2}\left(T_{i}^{2}+2m_{e}T_{i}\right)}
\ee
or scattering on earth
\be
\left(\frac{d\sigma}{dE_{\rm T}}\right)_{scalar,\rm{CR}}= \frac{\bar{\sigma}_e(\alpha^2m_e^2+m_{\phi}^2)^2 m_{\chi}m_{e}^{2}}{2 \mu_{\chi e}^{2}\left(2m_{e}T_{i}+m_{\phi}^{2}\right)^{2}\left(T_{\chi}^{2}+2m_{\chi}T_{\chi}\right)}
\ee

\subsection{Dark Photon Mediator}
Lastly, we examine the case of fermionic DM interacting with a kinetically mixed vector mediator.

The matrix element for incoming dark matter incident upon an electron which is scattered with a kinetic energy $E_T$ is
\bea
|\mathcal{M}_{vv}|^2 &=& \frac{8g_{v\chi}^2g_{v{\rm{CR}}}^2 (2m_\chi^2(m_e+T_i)^2 - m_\chi T_\chi((m_e+m_\chi)^2+2m_\chi T_i)+m_\chi^2 T_\chi^2)}{(2m_\chi T_\chi + m_v^2)^2} 
\\
\rightarrow |\mathcal{M}_{\rm{free}}|^2 &=& \frac{16g_\chi^2g_e^2m_\chi^2m_e^2}{(\alpha^2m_e^2 + m_v^2)^2}
\eea
We find the reference cross-section
\bea
\bar{\sigma}_{e,vv} = \frac{\mu_{\chi e}^2g_\chi^2g_e^2}{\pi(\alpha^2m_e^2+m_v^2)^2}
\eea
or equivalently, the mediator mass in terms of $\bar{\sigma}_e$
\bea
m_v^2 = \frac{\mu_{\chi e}g_\chi g_e }{\sqrt{\pi\bar{\sigma}_e}} - \alpha^2m_e^2
\eea
If we want to recast the reference cross-section in terms of a kinetically mixed dark photon, $A'$, with mixing parameter $\epsilon$, we can make the substitution $g_e^2 \rightarrow 4\pi\alpha\epsilon^2$ to arrive at the reference cross-section
\bea
\bar{\sigma}_{e,vv} = \frac{4\mu_{\chi e}^2g_\chi^2\alpha\epsilon^2}{\pi(\alpha^2m_e^2+m_{A'}^2)^2}
\eea

This can be applied to the differential cross-section for upscattering
\bea
\left(\frac{d\sigma}{dT_\chi}\right)_{vv,\rm{CR}} = g_{\chi}^2g_{e}^2\frac{(2m_\chi(m_e+T_i)^2 - T_\chi((m_e+m_\chi)^2+2m_\chi T_i)+m_\chi T_\chi^2)}{4\pi(2m_\chi T_\chi + m_v^2)^2(T_i^2 + 2m_eT_i)}
\eea
or scattering on earth
\bea
\left(\frac{d\sigma}{dE_{\rm T}}\right)_{vv,\rm{T}} = g_{\chi}^2g_{e}^2\frac{(2m_e(m_\chi+T_\chi)^2 - E_T((m_\chi+m_e)^2+2m_e T_\chi)+m_e E_T^2)}{4\pi(2m_e E_T + m_v^2)^2(T_\chi^2 + 2m_\chi T_\chi)}
\eea

\begin{figure*}[t!]
\includegraphics[angle=0,width=.32\textwidth]{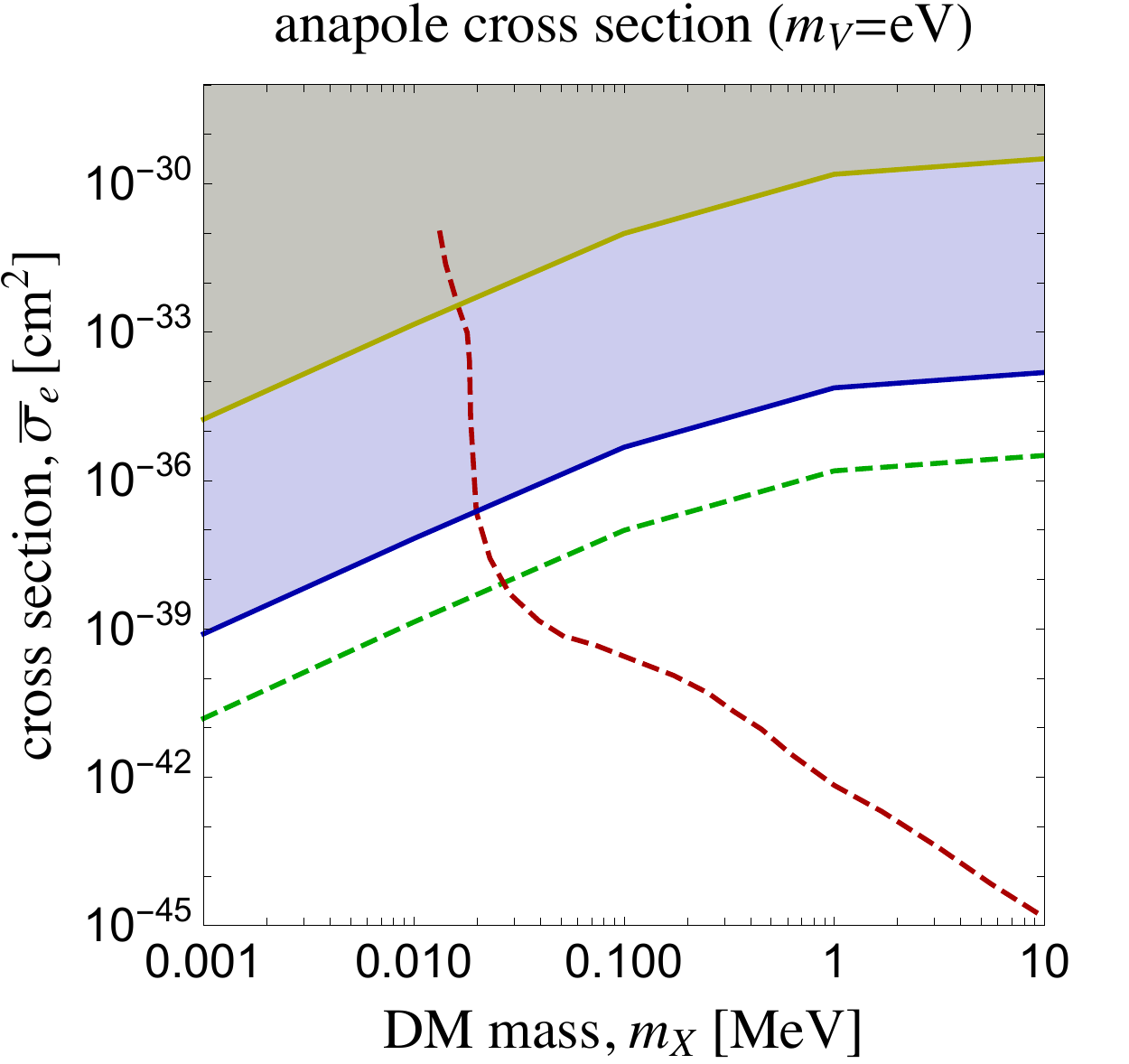}
\includegraphics[angle=0,width=.32\textwidth]{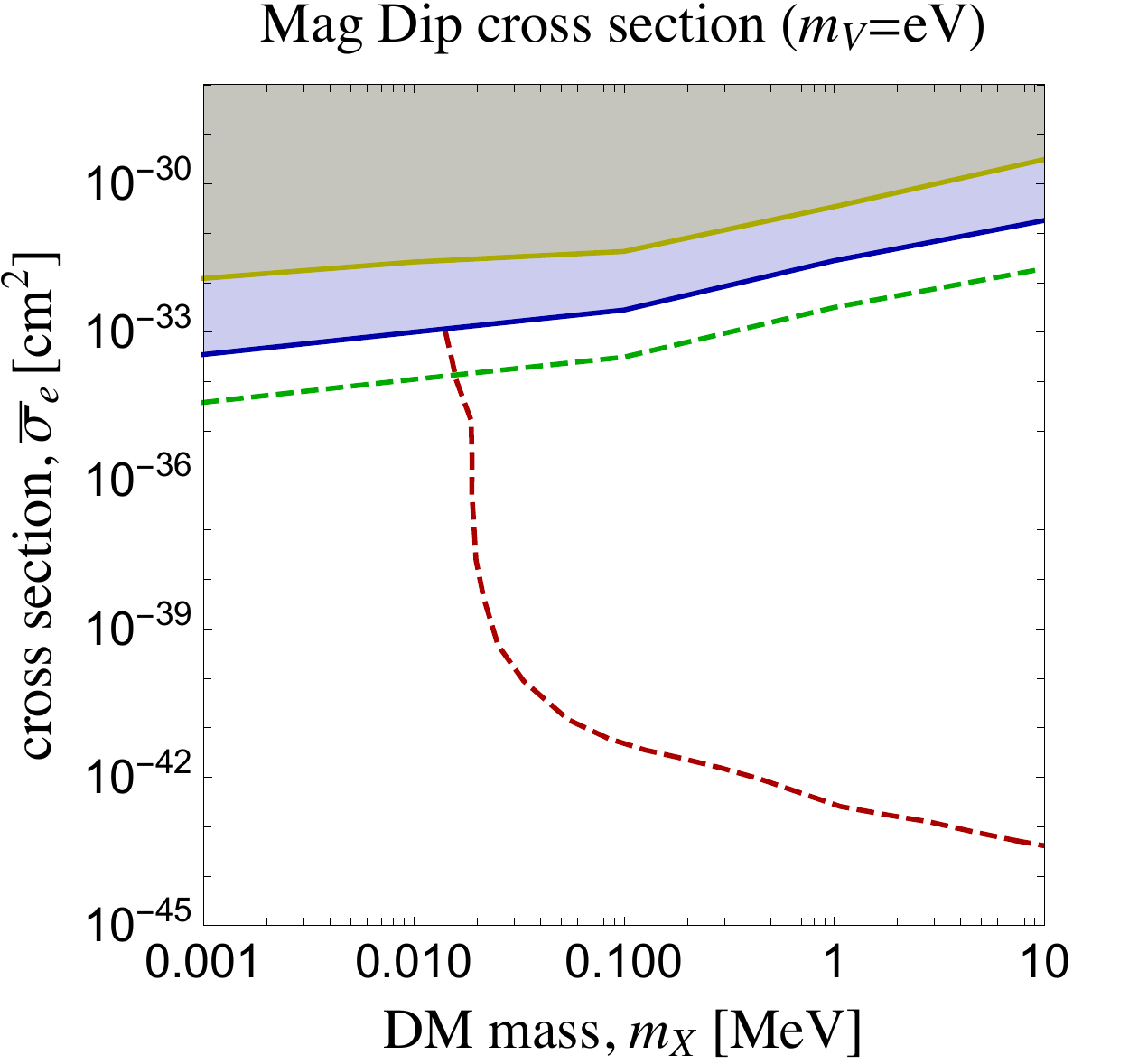}
\includegraphics[angle=0,width=.32\textwidth]{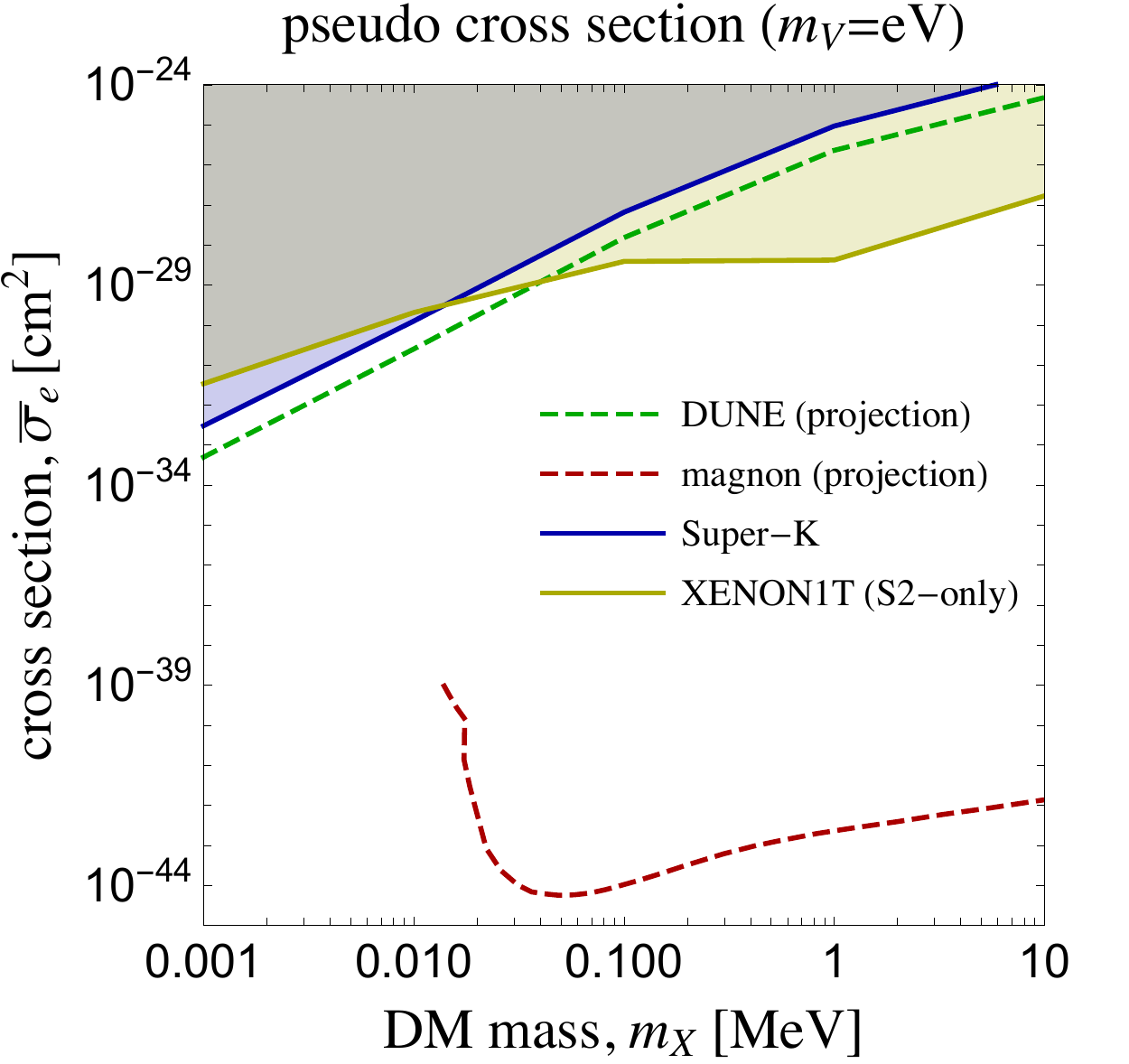}
\caption{Bounds from Super-K and XENON1T along with future magnon direct detection bounds~\cite{Trickle:2019ovy}, and our projected sensitivities for DUNE. }
\label{fig:fluxes}
\end{figure*}

\section{Experimental Sensitivities}
\label{sec:sensitivities}
For Super-K, we follow treatment in Ref.~\cite{Cappiello:2019qsw} by simply requiring that the bound not exceed Super-K's measured event rate in any bin. Typically, one expects the lowest energy bin to provide the strongest constraint since the signal rises at low energies. However, the signal detection efficiency drops near the low-energy threshold, so often the second bin provides the dominant constraint. In this analysis, we utilize the existing Super-K analysis~\cite{Bays:2011si} which was originally used to search for the Diffuse Supernova Background. We show the data points in Fig.~\ref{fig:SKcompare} along with predicted DM event rates in the magnetic dipole case. Note that the peak in the event rates arises from the signal efficiency dropping at low energies. We fold in the Super-K signal efficiency (Fig. 10 of Ref.~\cite{Bays:2011si}) in this search by convolving it with the theoretical event rate. For our Hyper-K estimates, we simply scale up this Super-K search by the corresponding detector mass.

As suggested in~\cite{Cappiello:2019qsw} it is possible that the upcoming gadolinium doping will allow for background reduction, thereby providing a pathway in the future for Super-K to improve on these constraints. Further, as done in~\cite{Ema:2018bih}, exploiting the directional sensitivity of Super-K could allow for some improvement in the signal-to-background ratio.

For XENON1T we use the S2-only analysis~\cite{Aprile:2019xxb} to explore the impact of a low threshold on the sensitivity. Following~\cite{Aprile:2019xxb} we require that no more than 13 events occur in the region 0.3 -3 keV$_{ee}$. 

In our DUNE projections, we follow~\cite{Acciarri:2015uup,Necib:2016aez}. For the 40 kton far detector we take $1.08 \times 10^{34}$ electron targets, with a 30 MeV detector threshold and assume a 10 year exposure. Background reduction will be crucial for this search. The dominant background at these energies arises from atmospheric neutrinos, via $\nu_{e} + n \longrightarrow e^{-} + p$, as well as, $\bar{\nu}_{e} + p \longrightarrow n+ e^{+}$. Given that these backgrounds induce some hadronic activity one can use such signals as a veto on this background. In~\cite{Necib:2016aez}, this background was carefully simulated and it was found that roughly $\sim 1/3$ of this background could passes the veto on hadronic activity. We will assume this level of background in the 90$\%$ CL sensitivity we project for DUNE. Note that this conservatively assumes that all neutrons in the final state will escape detection. Although neutron tagging is challenging, some fraction will be detectable, implying that this important background can be reduced further with additional detector simulations.

\begin{figure*}[t!]
\includegraphics[angle=0,width=.40\textwidth]{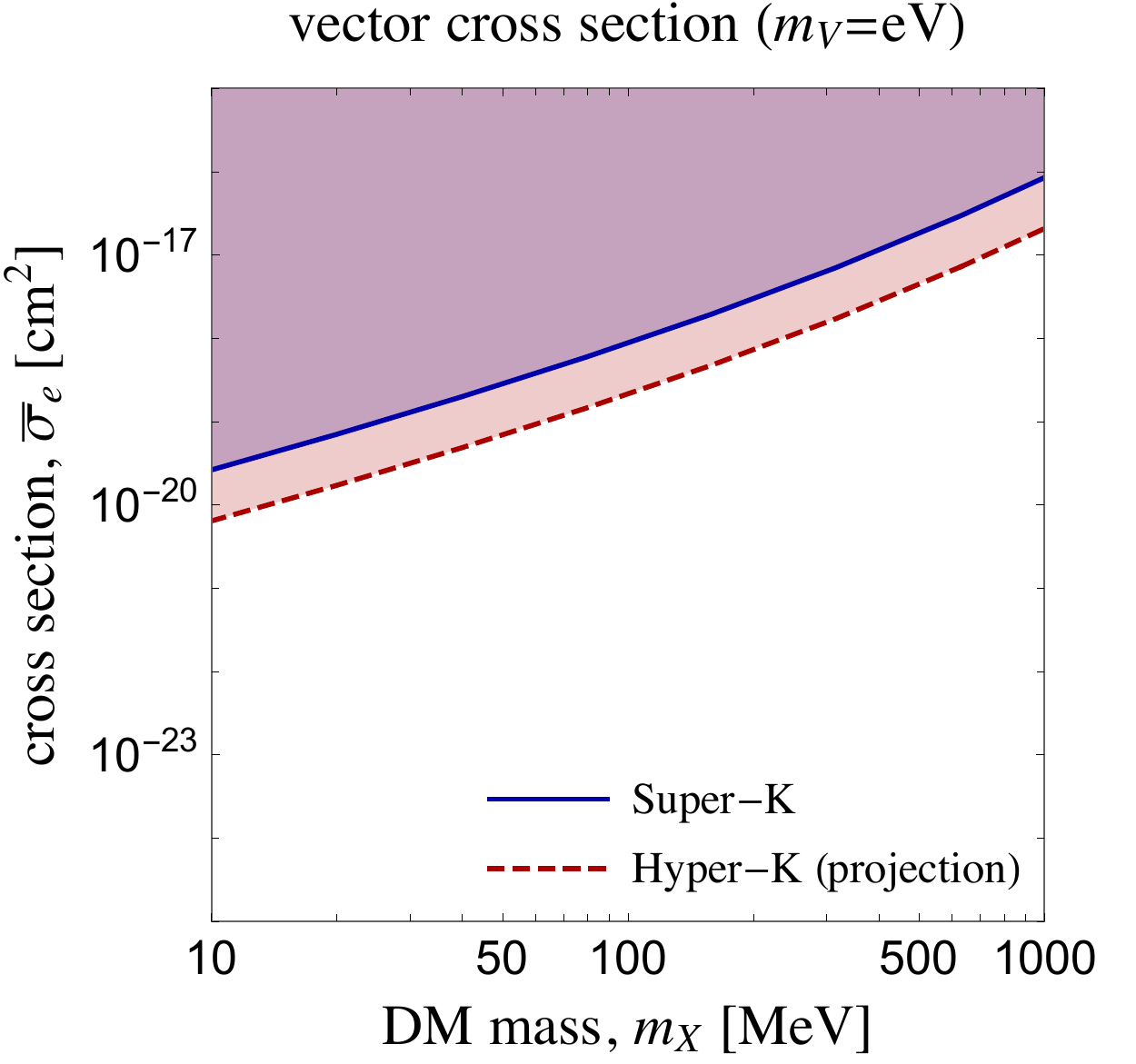}
\includegraphics[angle=0,width=.40\textwidth]{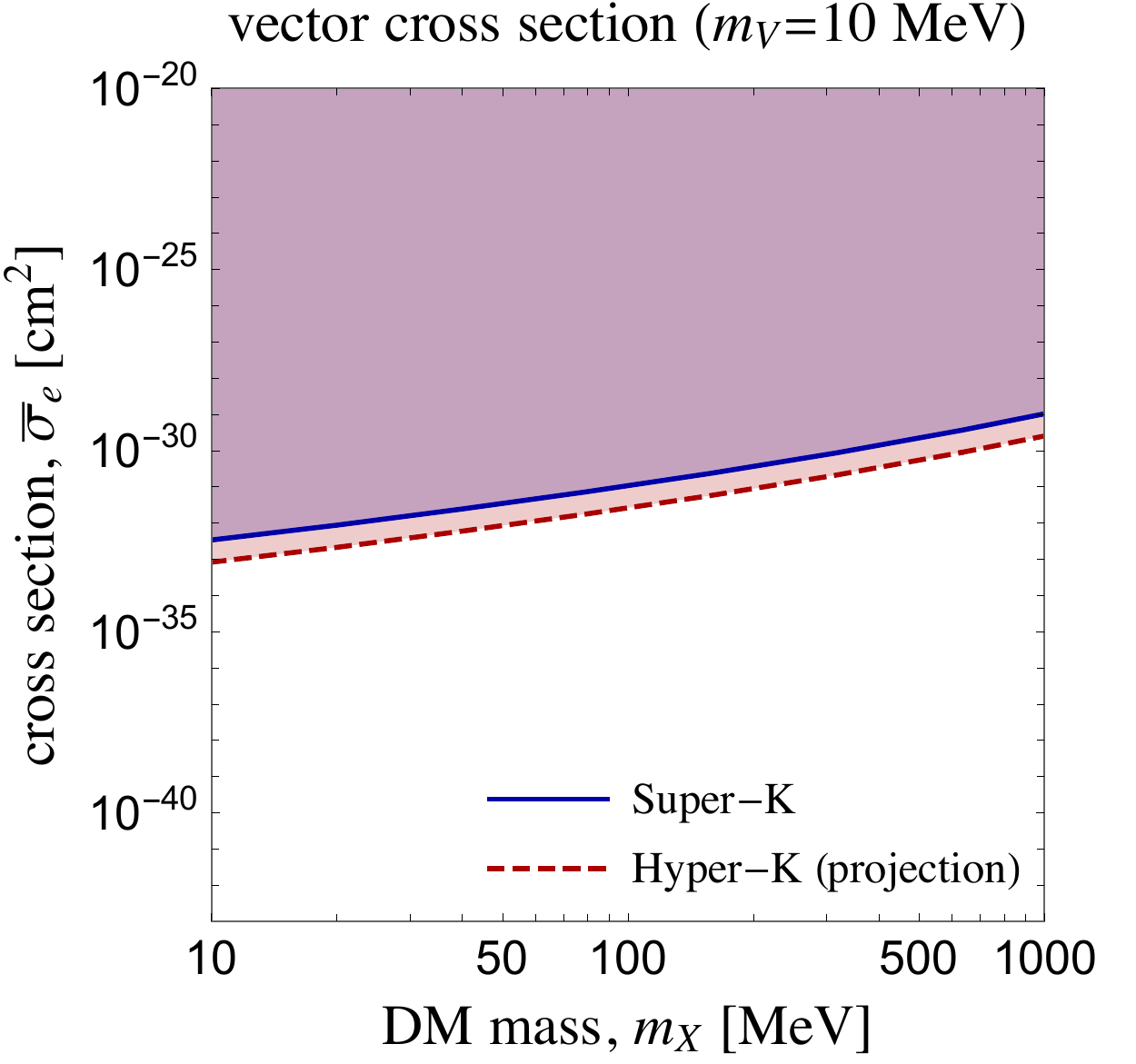}
\caption{Bounds from Super-K and Hyper-K for the case of a dark photon mediator~\cite{Trickle:2019ovy}. In the left panel we fix the dark photon mass to 1 eV, while in the right panel it is fixed to 10 MeV. See text for a discussion of additional bounds which are stronger.}
\label{fig:darkphoton}
\end{figure*}

Fig.~\ref{fig:fluxes} displays an interesting interplay of experimental sensitivities. At the highest DM masses, low-threshold direct detection experiments will be dominant. In this case we have illustrated this with the magnon-based experiments from~\cite{Trickle:2019ovy}. In the anapole and magnetic dipole interaction cases, the largest experimental exposure sets the strongest bounds. Thus in these cases we find that Super-K is dramatically stronger than the XENON1T bounds, and that DUNE will supersede Super-K. However, the steeply falling event rate of the pseudo-scalar model allows the XENON1T experiment to place stronger bounds than Super-K (right panel of Fig.~\ref{fig:fluxes}). Similar conclusions were recently observed in other models in~\cite{Bloch:2020uzh}.

Next we consider the vector mediator case, where we have assumed that the mediator couples to both electrons and protons. We compute bounds with the mediator mass fixed to both 1 eV and 10 MeV, which are shown, respectively, in the left and right panels of Fig.~\ref{fig:darkphoton}. First, we note that the sensitivity in the vector mediator case is increased compared to the other scenarios considered in this work since it benefits from the CR flux contributions from both protons and electrons.  When the mediator mass is light, our bounds are not competitive with the strong astrophysical constraints in this mass range. For example, at these low mediator masses the combination of White Dwarf, Red Giant, Supernovae, BBN, and XENON10 bounds disallow cross sections above the $\sim 10^{-33}~{\rm cm}^{2}$ level for keV to GeV DM masses~\cite{Knapen:2017xzo}. However, astrophysical constraints have large uncertainties and the bounds we derive are based on terrestrial data from the Super-K experiment and are in this sense complementary to the astrophysical constraints. Moreover, astrophysical constraints can be evaded in the context of chameleon models where the density of the astrophysical objects impact the mass of the mediators~\cite{Khoury:2003aq,Nelson_2008,Nelson:2008tn, DeRocco:2020xdt}. Lastly, note that fifth force searches (i.e. from precision tests of gravitational and van der Waals forces) can be very strong at low masses~\cite{Bordag:2001qi,Adelberger:2006dh,Adelberger:2009zz}, although for dark photons which kinetically mix with the photon these bounds are absent given the equal and opposite coupling of the dark photon to protons and electrons.

When the dark photon mass is heavier (10 MeV) we again find that our bounds are not competitive. However, in this case stronger bounds come from terrestrial experiments, including: E137~\cite{Bjorken:1988as}, Orsay~\cite{Davier:1989wz,Andreas:2012mt}, NA64~\cite{NA64:2019imj}, BaBar~\cite{Lees:2017lec}, and LSND~\cite{Kahn:2014sra,deNiverville:2011it,Batell:2009di}. Together these bounds preclude cross sections above the $\sim  10^{-37}~{\rm cm}^{2}$ level for MeV scale DM interacting via a 10 MeV dark photon~\cite{Essig:2015cda}.

One may worry that these cross sections are so large that they exceed the perturbativity bounds. Although this can be a concern in some cases~\cite{Digman:2019wdm}, we show in the Appendix that the couplings here remain fully perturbative. 

Lastly, we consider a leptophilic scalar mediator model with bounds shown in Fig.~\ref{fig:scalar}. Here we find that the low-threshold XENON1T analysis can provide stronger constraints than Super-K, due to the much lower energy sensitivity. These results are subject to additional astrophysical and cosmological constraints from a combination of Red Giant stars~\cite{Hardy:2016kme}, Horizontal Branch stars~\cite{Hardy:2016kme}, and BBN observations~\cite{Knapen:2017xzo}, as well as bounds from beam dump experiments~\cite{Liu:2016qwd}.


\begin{figure*}[t!]
\includegraphics[angle=0,width=.40\textwidth]{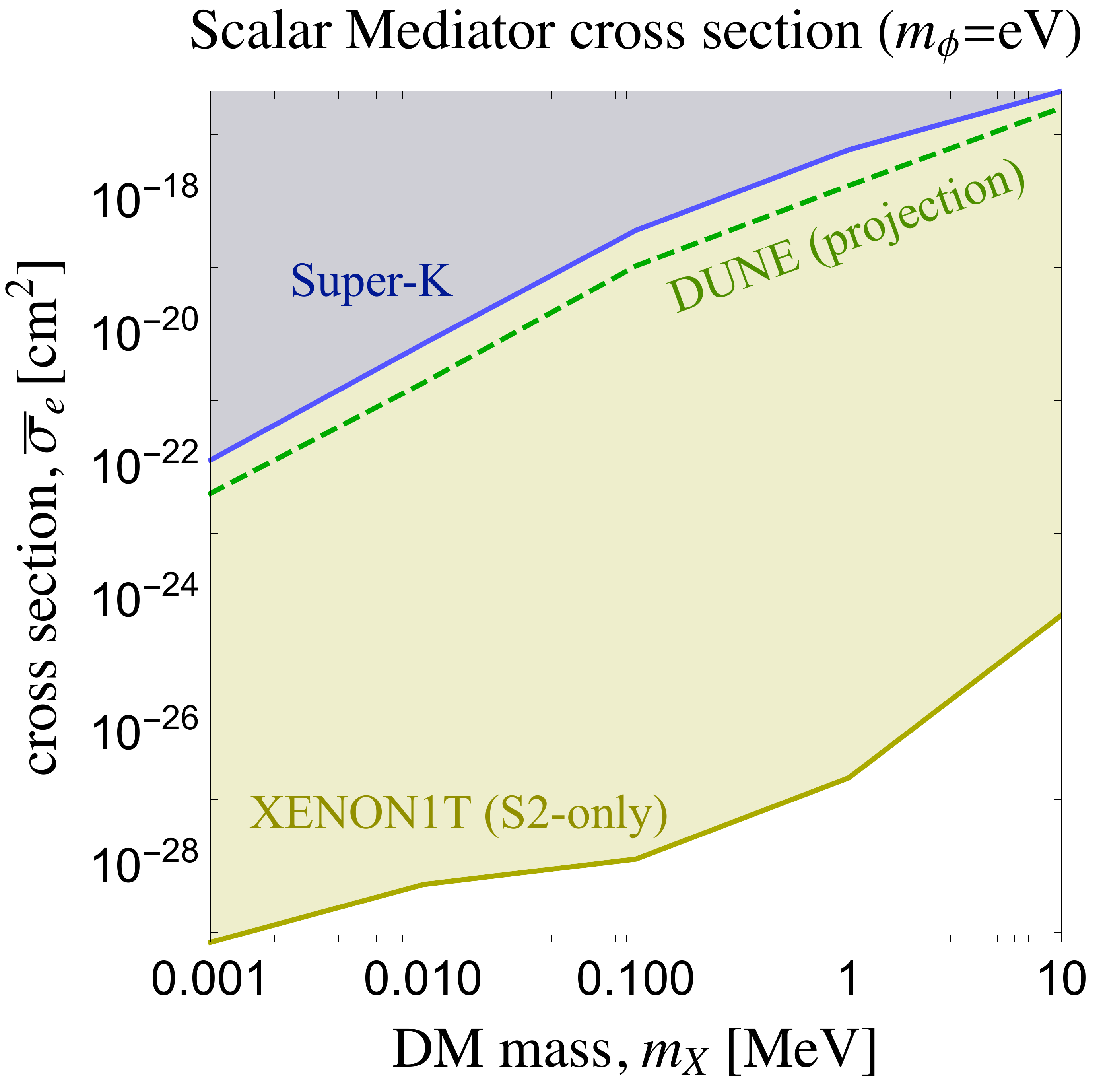}
\includegraphics[angle=0,width=.44\textwidth]{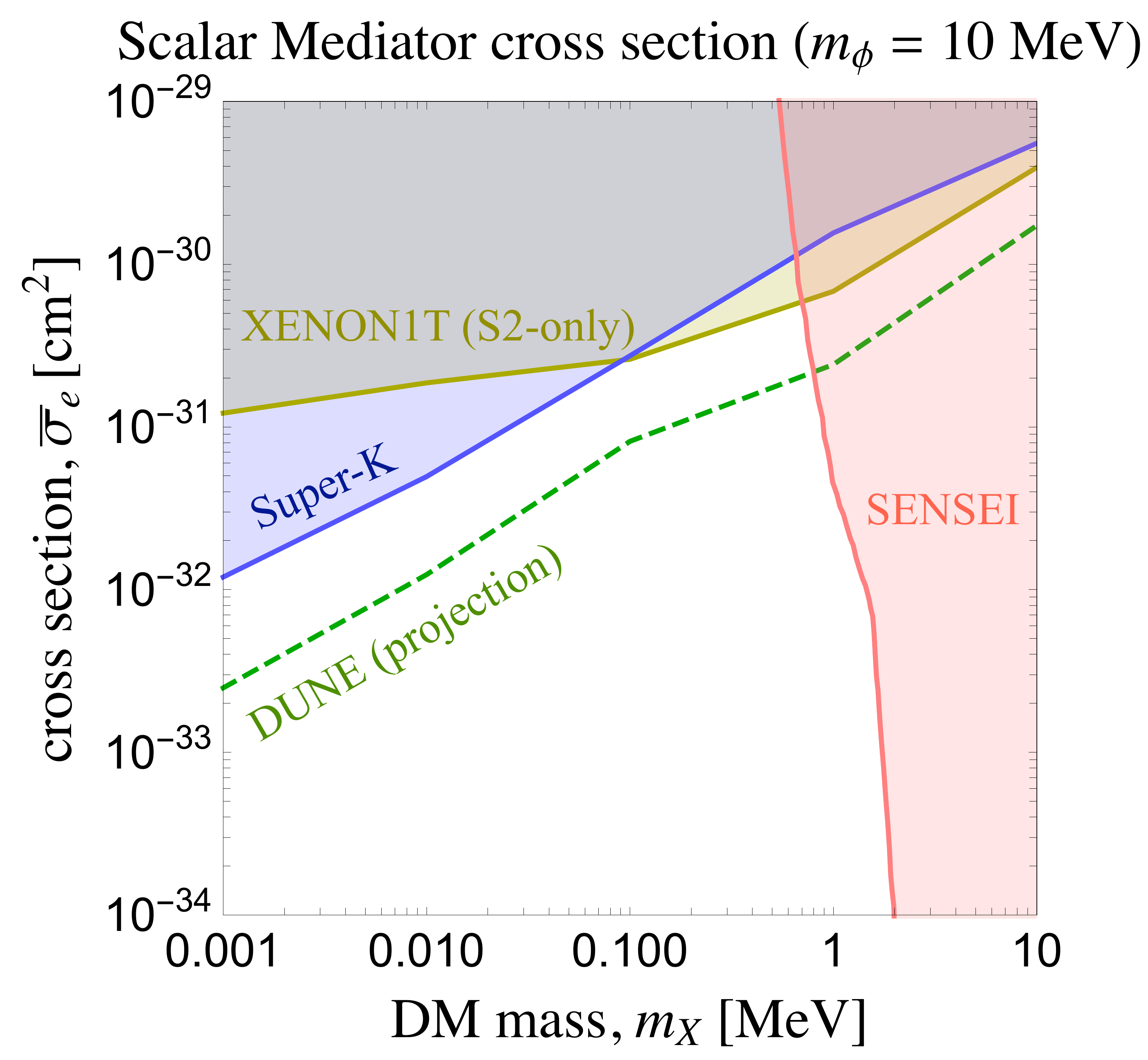}
\caption{Bounds from Super-K and XENON1T along with the projected DUNE sensitivity for the case of a leptophilic {scalar mediator}~\cite{Trickle:2019ovy}. In the left panel we fix the scalar mass to 1 eV, while in the right panel it is fixed to 10 MeV. We also include the SENSEI bounds~\cite{Barak:2020fql} shown in red in the right panel.}
\label{fig:scalar}
\end{figure*}

\section{Conclusions}
\label{sec:conclusions}

We have shown that present Super-K data produces leading constrains on DM-electron interactions at low 
DM masses. It is important to stress that these bounds are conservative given that they are only based on the relatively small exposure from the 2853 days used in the supernova relic search~\cite{Bays:2011si}. In addition to the improvements in exposure, one could also improve on the search strategy utilized here by background subtraction, spectral shape, and directional information. 

In the future we have found that DUNE and Hyper-K will be able to extend the reach significantly for the search for CR upscattered DM for a wide class of possible DM-electron interactions. Moreover low-threshold direct detection experiments display novel complementarity to the neutrino experiments, since they can be uniquely sensitive to those models with rates that fall steeply with energy.

Finally, we note that while BBN considerations can impose strong constraints on some of the models discussed here~\cite{Knapen:2017xzo,Krnjaic:2019dzc,Sabti:2019mhn}, these bounds also make assumptions which may be relaxed in models where DM has both neutrino and electron couplings~\cite{Escudero:2018mvt}.

\vspace{1cm}

{\bf \emph{Acknowledgements-  }}The work of B.D. is supported in part by the DOE Grant No. DE-SC0010813. The work of IMS and NTA is supported by the U.S. Department of Energy  under the award number DE-SC0020250. JLN is supported by the Australian Research Council. JBD acknowledges support from the National Science Foundation under Grant No. NSF PHY-1820801.


\newpage 
\appendix
\section{Large Cross Sections with small couplings}
Here we include for reference additional figures in Fig.~\ref{fig:coupling} showing the experimental sensitivity on the coupling. This illustrates that although the cross section bounds displayed in the main body of the text are fairly large, they are well below perturbativity bounds.

\begin{figure*}[t!]
\includegraphics[angle=0,width=.44\textwidth]{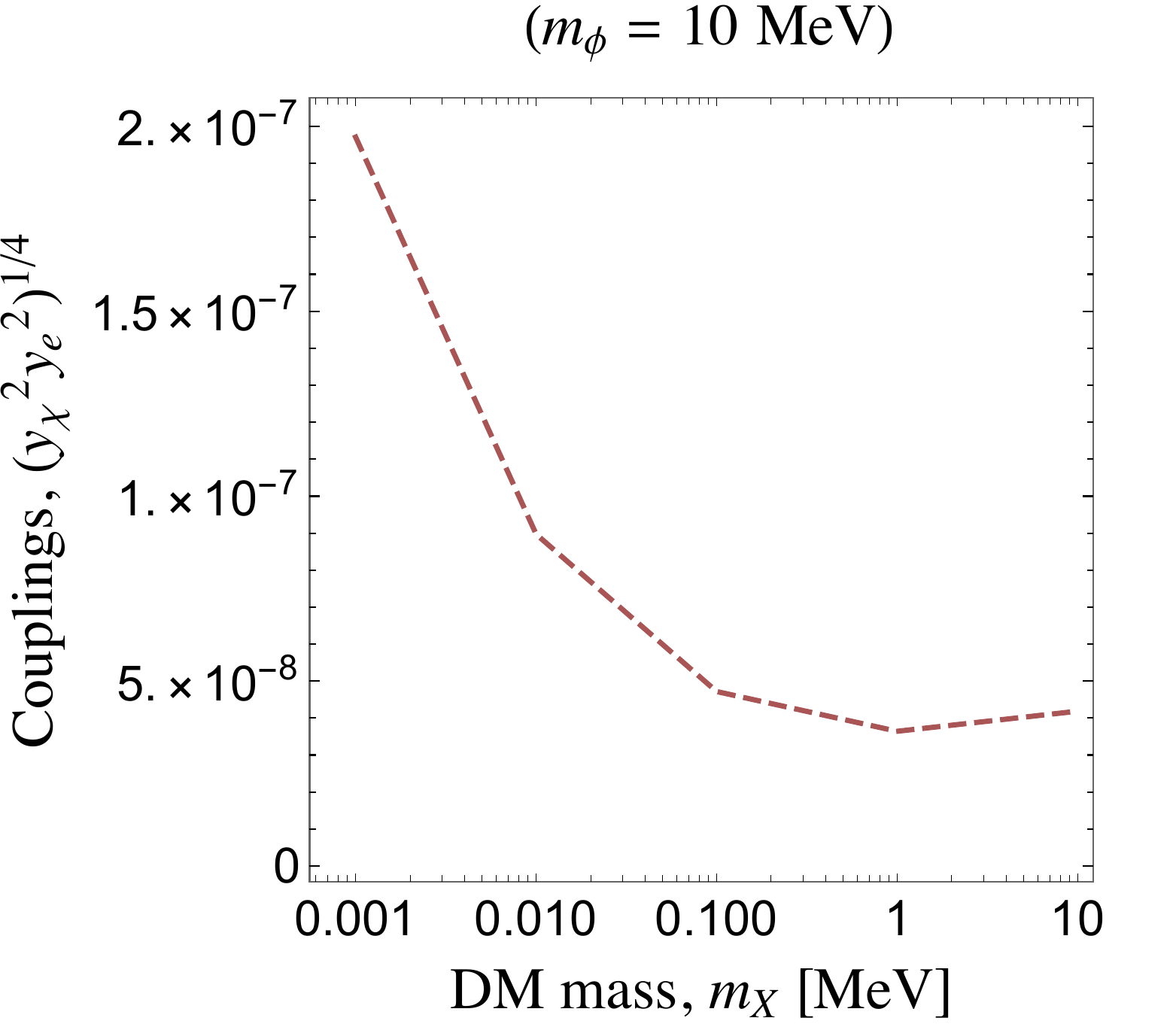}
\includegraphics[angle=0,width=.44\textwidth]{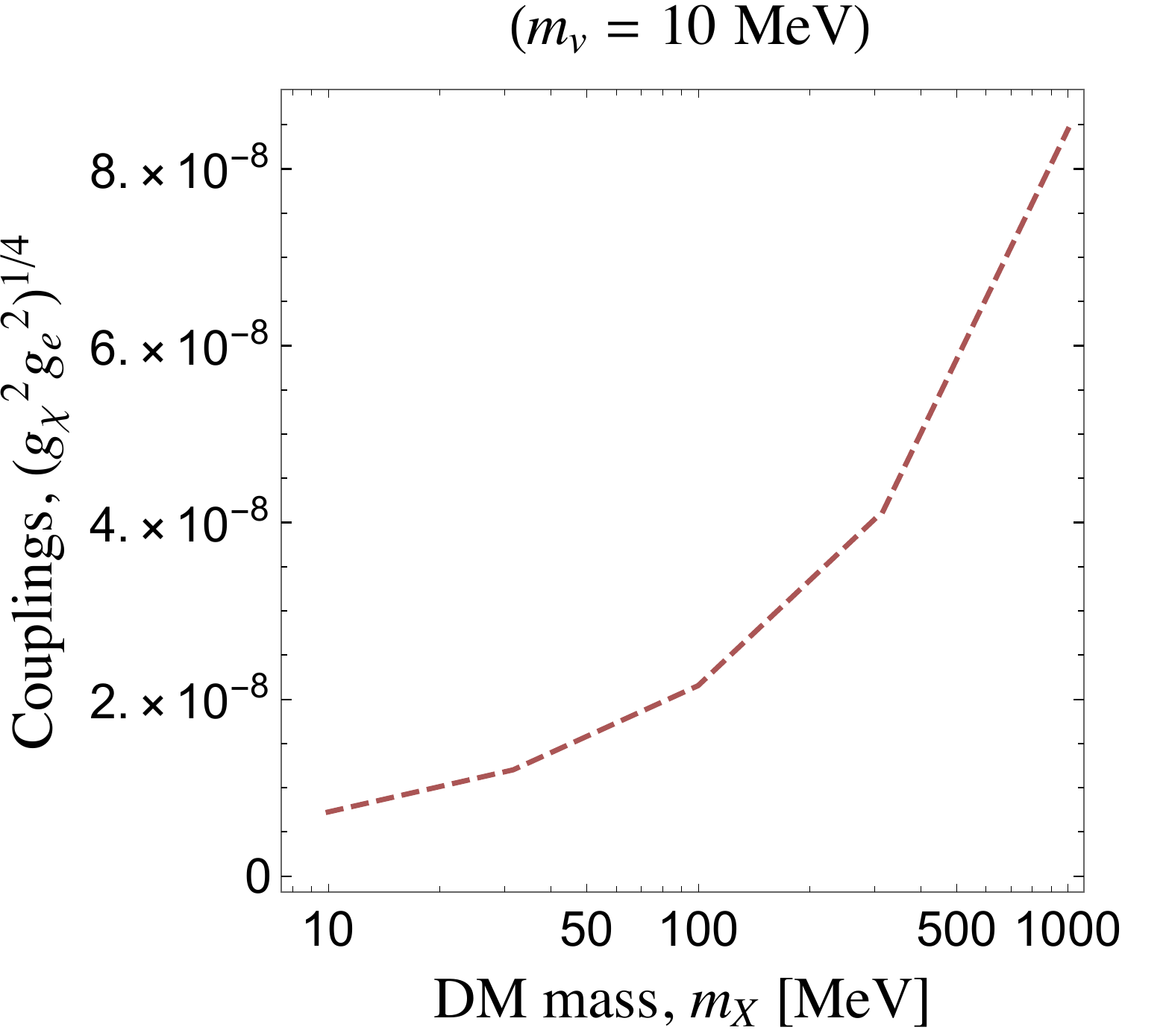}
\caption{We explicitly display bounds on the Lagrangian-level coupling based on Super-K data in the scalar (left) and vector mediator (right) models. This is an alternative choice compared with the cross section bounds in the main body of the text. }
\label{fig:coupling}
\end{figure*}

\bibliography{crdm}

\begin{thebibliography}{57}%
\makeatletter
\providecommand \@ifxundefined [1]{%
 \@ifx{#1\undefined}
}%
\providecommand \@ifnum [1]{%
 \ifnum #1\expandafter \@firstoftwo
 \else \expandafter \@secondoftwo
 \fi
}%
\providecommand \@ifx [1]{%
 \ifx #1\expandafter \@firstoftwo
 \else \expandafter \@secondoftwo
 \fi
}%
\providecommand \natexlab [1]{#1}%
\providecommand \enquote  [1]{``#1''}%
\providecommand \bibnamefont  [1]{#1}%
\providecommand \bibfnamefont [1]{#1}%
\providecommand \citenamefont [1]{#1}%
\providecommand \href@noop [0]{\@secondoftwo}%
\providecommand \href [0]{\begingroup \@sanitize@url \@href}%
\providecommand \@href[1]{\@@startlink{#1}\@@href}%
\providecommand \@@href[1]{\endgroup#1\@@endlink}%
\providecommand \@sanitize@url [0]{\catcode `\\12\catcode `\$12\catcode
  `\&12\catcode `\#12\catcode `\^12\catcode `\_12\catcode `\%12\relax}%
\providecommand \@@startlink[1]{}%
\providecommand \@@endlink[0]{}%
\providecommand \url  [0]{\begingroup\@sanitize@url \@url }%
\providecommand \@url [1]{\endgroup\@href {#1}{\urlprefix }}%
\providecommand \urlprefix  [0]{URL }%
\providecommand \Eprint [0]{\href }%
\providecommand \doibase [0]{http://dx.doi.org/}%
\providecommand \selectlanguage [0]{\@gobble}%
\providecommand \bibinfo  [0]{\@secondoftwo}%
\providecommand \bibfield  [0]{\@secondoftwo}%
\providecommand \translation [1]{[#1]}%
\providecommand \BibitemOpen [0]{}%
\providecommand \bibitemStop [0]{}%
\providecommand \bibitemNoStop [0]{.\EOS\space}%
\providecommand \EOS [0]{\spacefactor3000\relax}%
\providecommand \BibitemShut  [1]{\csname bibitem#1\endcsname}%
\let\auto@bib@innerbib\@empty
\bibitem [{\citenamefont {Goodman}\ and\ \citenamefont
  {Witten}(1985)}]{Goodman:1984dc}%
  \BibitemOpen
  \bibfield  {author} {\bibinfo {author} {\bibfnamefont {M.~W.}\ \bibnamefont
  {Goodman}}\ and\ \bibinfo {author} {\bibfnamefont {E.}~\bibnamefont
  {Witten}},\ }\href {\doibase 10.1103/PhysRevD.31.3059} {\bibfield  {journal}
  {\bibinfo  {journal} {Phys. Rev.}\ }\textbf {\bibinfo {volume} {D31}},\
  \bibinfo {pages} {3059} (\bibinfo {year} {1985})},\ \bibinfo {note}
  {[,325(1984)]}\BibitemShut {NoStop}%
\bibitem [{\citenamefont {Ema}\ \emph {et~al.}(2018)\citenamefont {Ema},
  \citenamefont {Sala},\ and\ \citenamefont {Sato}}]{Ema:2018bih}%
  \BibitemOpen
  \bibfield  {author} {\bibinfo {author} {\bibfnamefont {Y.}~\bibnamefont
  {Ema}}, \bibinfo {author} {\bibfnamefont {F.}~\bibnamefont {Sala}}, \ and\
  \bibinfo {author} {\bibfnamefont {R.}~\bibnamefont {Sato}},\ }\href@noop {}
  {\  (\bibinfo {year} {2018})},\ \Eprint {http://arxiv.org/abs/1811.00520}
  {arXiv:1811.00520 [hep-ph]} \BibitemShut {NoStop}%
\bibitem [{\citenamefont {Bringmann}\ and\ \citenamefont
  {Pospelov}(2018)}]{Bringmann:2018cvk}%
  \BibitemOpen
  \bibfield  {author} {\bibinfo {author} {\bibfnamefont {T.}~\bibnamefont
  {Bringmann}}\ and\ \bibinfo {author} {\bibfnamefont {M.}~\bibnamefont
  {Pospelov}},\ }\href@noop {} {\  (\bibinfo {year} {2018})},\ \Eprint
  {http://arxiv.org/abs/1810.10543} {arXiv:1810.10543 [hep-ph]} \BibitemShut
  {NoStop}%
\bibitem [{\citenamefont {Dent}\ \emph {et~al.}(2020)\citenamefont {Dent},
  \citenamefont {Dutta}, \citenamefont {Newstead},\ and\ \citenamefont
  {Shoemaker}}]{Dent:2019krz}%
  \BibitemOpen
  \bibfield  {author} {\bibinfo {author} {\bibfnamefont {J.~B.}\ \bibnamefont
  {Dent}}, \bibinfo {author} {\bibfnamefont {B.}~\bibnamefont {Dutta}},
  \bibinfo {author} {\bibfnamefont {J.~L.}\ \bibnamefont {Newstead}}, \ and\
  \bibinfo {author} {\bibfnamefont {I.~M.}\ \bibnamefont {Shoemaker}},\ }\href
  {\doibase 10.1103/PhysRevD.101.116007} {\bibfield  {journal} {\bibinfo
  {journal} {Phys. Rev. D}\ }\textbf {\bibinfo {volume} {101}},\ \bibinfo
  {pages} {116007} (\bibinfo {year} {2020})},\ \Eprint
  {http://arxiv.org/abs/1907.03782} {arXiv:1907.03782 [hep-ph]} \BibitemShut
  {NoStop}%
\bibitem [{\citenamefont {Cappiello}\ and\ \citenamefont
  {Beacom}(2019)}]{Cappiello:2019qsw}%
  \BibitemOpen
  \bibfield  {author} {\bibinfo {author} {\bibfnamefont {C.}~\bibnamefont
  {Cappiello}}\ and\ \bibinfo {author} {\bibfnamefont {J.~F.}\ \bibnamefont
  {Beacom}},\ }\href {\doibase 10.1103/PhysRevD.100.103011} {\bibfield
  {journal} {\bibinfo  {journal} {Phys. Rev. D}\ }\textbf {\bibinfo {volume}
  {100}},\ \bibinfo {pages} {103011} (\bibinfo {year} {2019})},\ \Eprint
  {http://arxiv.org/abs/1906.11283} {arXiv:1906.11283 [hep-ph]} \BibitemShut
  {NoStop}%
\bibitem [{\citenamefont {Cappiello}\ \emph {et~al.}(2019)\citenamefont
  {Cappiello}, \citenamefont {Ng},\ and\ \citenamefont
  {Beacom}}]{Cappiello:2018hsu}%
  \BibitemOpen
  \bibfield  {author} {\bibinfo {author} {\bibfnamefont {C.~V.}\ \bibnamefont
  {Cappiello}}, \bibinfo {author} {\bibfnamefont {K.~C.~Y.}\ \bibnamefont
  {Ng}}, \ and\ \bibinfo {author} {\bibfnamefont {J.~F.}\ \bibnamefont
  {Beacom}},\ }\href {\doibase 10.1103/PhysRevD.99.063004} {\bibfield
  {journal} {\bibinfo  {journal} {Phys. Rev.}\ }\textbf {\bibinfo {volume}
  {D99}},\ \bibinfo {pages} {063004} (\bibinfo {year} {2019})},\ \Eprint
  {http://arxiv.org/abs/1810.07705} {arXiv:1810.07705 [hep-ph]} \BibitemShut
  {NoStop}%
\bibitem [{\citenamefont {Ibe}\ \emph {et~al.}(2018)\citenamefont {Ibe},
  \citenamefont {Nakano}, \citenamefont {Shoji},\ and\ \citenamefont
  {Suzuki}}]{Ibe:2017yqa}%
  \BibitemOpen
  \bibfield  {author} {\bibinfo {author} {\bibfnamefont {M.}~\bibnamefont
  {Ibe}}, \bibinfo {author} {\bibfnamefont {W.}~\bibnamefont {Nakano}},
  \bibinfo {author} {\bibfnamefont {Y.}~\bibnamefont {Shoji}}, \ and\ \bibinfo
  {author} {\bibfnamefont {K.}~\bibnamefont {Suzuki}},\ }\href {\doibase
  10.1007/JHEP03(2018)194} {\bibfield  {journal} {\bibinfo  {journal} {JHEP}\
  }\textbf {\bibinfo {volume} {03}},\ \bibinfo {pages} {194} (\bibinfo {year}
  {2018})},\ \Eprint {http://arxiv.org/abs/1707.07258} {arXiv:1707.07258
  [hep-ph]} \BibitemShut {NoStop}%
\bibitem [{\citenamefont {Dolan}\ \emph {et~al.}(2018)\citenamefont {Dolan},
  \citenamefont {Kahlhoefer},\ and\ \citenamefont {McCabe}}]{Dolan:2017xbu}%
  \BibitemOpen
  \bibfield  {author} {\bibinfo {author} {\bibfnamefont {M.~J.}\ \bibnamefont
  {Dolan}}, \bibinfo {author} {\bibfnamefont {F.}~\bibnamefont {Kahlhoefer}}, \
  and\ \bibinfo {author} {\bibfnamefont {C.}~\bibnamefont {McCabe}},\ }\href
  {\doibase 10.1103/PhysRevLett.121.101801} {\bibfield  {journal} {\bibinfo
  {journal} {Phys. Rev. Lett.}\ }\textbf {\bibinfo {volume} {121}},\ \bibinfo
  {pages} {101801} (\bibinfo {year} {2018})},\ \Eprint
  {http://arxiv.org/abs/1711.09906} {arXiv:1711.09906 [hep-ph]} \BibitemShut
  {NoStop}%
\bibitem [{\citenamefont {Akerib}\ \emph {et~al.}(2019)\citenamefont {Akerib}
  \emph {et~al.}}]{Akerib:2018hck}%
  \BibitemOpen
  \bibfield  {author} {\bibinfo {author} {\bibfnamefont {D.~S.}\ \bibnamefont
  {Akerib}} \emph {et~al.} (\bibinfo {collaboration} {LUX}),\ }\href {\doibase
  10.1103/PhysRevLett.122.131301} {\bibfield  {journal} {\bibinfo  {journal}
  {Phys. Rev. Lett.}\ }\textbf {\bibinfo {volume} {122}},\ \bibinfo {pages}
  {131301} (\bibinfo {year} {2019})},\ \Eprint
  {http://arxiv.org/abs/1811.11241} {arXiv:1811.11241 [astro-ph.CO]}
  \BibitemShut {NoStop}%
\bibitem [{\citenamefont {Armengaud}\ \emph {et~al.}(2019)\citenamefont
  {Armengaud} \emph {et~al.}}]{Armengaud:2019kfj}%
  \BibitemOpen
  \bibfield  {author} {\bibinfo {author} {\bibfnamefont {E.}~\bibnamefont
  {Armengaud}} \emph {et~al.} (\bibinfo {collaboration} {EDELWEISS}),\ }\href
  {\doibase 10.1103/PhysRevD.99.082003} {\bibfield  {journal} {\bibinfo
  {journal} {Phys. Rev.}\ }\textbf {\bibinfo {volume} {D99}},\ \bibinfo {pages}
  {082003} (\bibinfo {year} {2019})},\ \Eprint
  {http://arxiv.org/abs/1901.03588} {arXiv:1901.03588 [astro-ph.GA]}
  \BibitemShut {NoStop}%
\bibitem [{\citenamefont {Bell}\ \emph {et~al.}(2019)\citenamefont {Bell},
  \citenamefont {Dent}, \citenamefont {Newstead}, \citenamefont {Sabharwale},\
  and\ \citenamefont {Weiler}}]{Bell:2019egg}%
  \BibitemOpen
  \bibfield  {author} {\bibinfo {author} {\bibfnamefont {N.~F.}\ \bibnamefont
  {Bell}}, \bibinfo {author} {\bibfnamefont {J.~B.}\ \bibnamefont {Dent}},
  \bibinfo {author} {\bibfnamefont {J.~L.}\ \bibnamefont {Newstead}}, \bibinfo
  {author} {\bibfnamefont {S.}~\bibnamefont {Sabharwale}}, \ and\ \bibinfo
  {author} {\bibfnamefont {T.~J.}\ \bibnamefont {Weiler}},\ }\href@noop {} {\
  (\bibinfo {year} {2019})},\ \Eprint {http://arxiv.org/abs/1905.00046}
  {arXiv:1905.00046 [hep-ph]} \BibitemShut {NoStop}%
\bibitem [{\citenamefont {Liu}\ \emph {et~al.}(2019)\citenamefont {Liu} \emph
  {et~al.}}]{Liu:2019kzq}%
  \BibitemOpen
  \bibfield  {author} {\bibinfo {author} {\bibfnamefont {Z.~Z.}\ \bibnamefont
  {Liu}} \emph {et~al.} (\bibinfo {collaboration} {CDEX}),\ }\href@noop {} {\
  (\bibinfo {year} {2019})},\ \Eprint {http://arxiv.org/abs/1905.00354}
  {arXiv:1905.00354 [hep-ex]} \BibitemShut {NoStop}%
\bibitem [{\citenamefont {Kouvaris}\ and\ \citenamefont
  {Pradler}(2017)}]{Kouvaris:2016afs}%
  \BibitemOpen
  \bibfield  {author} {\bibinfo {author} {\bibfnamefont {C.}~\bibnamefont
  {Kouvaris}}\ and\ \bibinfo {author} {\bibfnamefont {J.}~\bibnamefont
  {Pradler}},\ }\href {\doibase 10.1103/PhysRevLett.118.031803} {\bibfield
  {journal} {\bibinfo  {journal} {Phys. Rev. Lett.}\ }\textbf {\bibinfo
  {volume} {118}},\ \bibinfo {pages} {031803} (\bibinfo {year} {2017})},\
  \Eprint {http://arxiv.org/abs/1607.01789} {arXiv:1607.01789 [hep-ph]}
  \BibitemShut {NoStop}%
\bibitem [{\citenamefont {Gluscevic}\ and\ \citenamefont
  {Boddy}(2018)}]{Gluscevic:2017ywp}%
  \BibitemOpen
  \bibfield  {author} {\bibinfo {author} {\bibfnamefont {V.}~\bibnamefont
  {Gluscevic}}\ and\ \bibinfo {author} {\bibfnamefont {K.~K.}\ \bibnamefont
  {Boddy}},\ }\href {\doibase 10.1103/PhysRevLett.121.081301} {\bibfield
  {journal} {\bibinfo  {journal} {Phys. Rev. Lett.}\ }\textbf {\bibinfo
  {volume} {121}},\ \bibinfo {pages} {081301} (\bibinfo {year} {2018})},\
  \Eprint {http://arxiv.org/abs/1712.07133} {arXiv:1712.07133 [astro-ph.CO]}
  \BibitemShut {NoStop}%
\bibitem [{\citenamefont {Boddy}\ and\ \citenamefont
  {Gluscevic}(2018)}]{Boddy:2018kfv}%
  \BibitemOpen
  \bibfield  {author} {\bibinfo {author} {\bibfnamefont {K.~K.}\ \bibnamefont
  {Boddy}}\ and\ \bibinfo {author} {\bibfnamefont {V.}~\bibnamefont
  {Gluscevic}},\ }\href {\doibase 10.1103/PhysRevD.98.083510} {\bibfield
  {journal} {\bibinfo  {journal} {Phys. Rev.}\ }\textbf {\bibinfo {volume}
  {D98}},\ \bibinfo {pages} {083510} (\bibinfo {year} {2018})},\ \Eprint
  {http://arxiv.org/abs/1801.08609} {arXiv:1801.08609 [astro-ph.CO]}
  \BibitemShut {NoStop}%
\bibitem [{\citenamefont {Boddy}\ \emph {et~al.}(2018)\citenamefont {Boddy},
  \citenamefont {Gluscevic}, \citenamefont {Poulin}, \citenamefont {Kovetz},
  \citenamefont {Kamionkowski},\ and\ \citenamefont {Barkana}}]{Boddy:2018wzy}%
  \BibitemOpen
  \bibfield  {author} {\bibinfo {author} {\bibfnamefont {K.~K.}\ \bibnamefont
  {Boddy}}, \bibinfo {author} {\bibfnamefont {V.}~\bibnamefont {Gluscevic}},
  \bibinfo {author} {\bibfnamefont {V.}~\bibnamefont {Poulin}}, \bibinfo
  {author} {\bibfnamefont {E.~D.}\ \bibnamefont {Kovetz}}, \bibinfo {author}
  {\bibfnamefont {M.}~\bibnamefont {Kamionkowski}}, \ and\ \bibinfo {author}
  {\bibfnamefont {R.}~\bibnamefont {Barkana}},\ }\href {\doibase
  10.1103/PhysRevD.98.123506} {\bibfield  {journal} {\bibinfo  {journal} {Phys.
  Rev.}\ }\textbf {\bibinfo {volume} {D98}},\ \bibinfo {pages} {123506}
  (\bibinfo {year} {2018})},\ \Eprint {http://arxiv.org/abs/1808.00001}
  {arXiv:1808.00001 [astro-ph.CO]} \BibitemShut {NoStop}%
\bibitem [{\citenamefont {Su}\ \emph {et~al.}(2020)\citenamefont {Su},
  \citenamefont {Wang}, \citenamefont {Wu}, \citenamefont {Yang},\ and\
  \citenamefont {Zhu}}]{Su:2020zny}%
  \BibitemOpen
  \bibfield  {author} {\bibinfo {author} {\bibfnamefont {L.}~\bibnamefont
  {Su}}, \bibinfo {author} {\bibfnamefont {W.}~\bibnamefont {Wang}}, \bibinfo
  {author} {\bibfnamefont {L.}~\bibnamefont {Wu}}, \bibinfo {author}
  {\bibfnamefont {J.~M.}\ \bibnamefont {Yang}}, \ and\ \bibinfo {author}
  {\bibfnamefont {B.}~\bibnamefont {Zhu}},\ }\href@noop {} {\  (\bibinfo {year}
  {2020})},\ \Eprint {http://arxiv.org/abs/2006.11837} {arXiv:2006.11837
  [hep-ph]} \BibitemShut {NoStop}%
\bibitem [{\citenamefont {Cao}\ \emph {et~al.}(2020)\citenamefont {Cao},
  \citenamefont {Ding},\ and\ \citenamefont {Xiang}}]{Cao:2020bwd}%
  \BibitemOpen
  \bibfield  {author} {\bibinfo {author} {\bibfnamefont {Q.-H.}\ \bibnamefont
  {Cao}}, \bibinfo {author} {\bibfnamefont {R.}~\bibnamefont {Ding}}, \ and\
  \bibinfo {author} {\bibfnamefont {Q.-F.}\ \bibnamefont {Xiang}},\ }\href@noop
  {} {\  (\bibinfo {year} {2020})},\ \Eprint {http://arxiv.org/abs/2006.12767}
  {arXiv:2006.12767 [hep-ph]} \BibitemShut {NoStop}%
\bibitem [{\citenamefont {Bloch}\ \emph {et~al.}(2020)\citenamefont {Bloch},
  \citenamefont {Caputo}, \citenamefont {Essig}, \citenamefont {Redigolo},
  \citenamefont {Sholapurkar},\ and\ \citenamefont {Volansky}}]{Bloch:2020uzh}%
  \BibitemOpen
  \bibfield  {author} {\bibinfo {author} {\bibfnamefont {I.~M.}\ \bibnamefont
  {Bloch}}, \bibinfo {author} {\bibfnamefont {A.}~\bibnamefont {Caputo}},
  \bibinfo {author} {\bibfnamefont {R.}~\bibnamefont {Essig}}, \bibinfo
  {author} {\bibfnamefont {D.}~\bibnamefont {Redigolo}}, \bibinfo {author}
  {\bibfnamefont {M.}~\bibnamefont {Sholapurkar}}, \ and\ \bibinfo {author}
  {\bibfnamefont {T.}~\bibnamefont {Volansky}},\ }\href@noop {} {\  (\bibinfo
  {year} {2020})},\ \Eprint {http://arxiv.org/abs/2006.14521} {arXiv:2006.14521
  [hep-ph]} \BibitemShut {NoStop}%
\bibitem [{\citenamefont {Zhang}\ \emph {et~al.}(2020)\citenamefont {Zhang},
  \citenamefont {Lei},\ and\ \citenamefont {Tang}}]{Zhang:2020htl}%
  \BibitemOpen
  \bibfield  {author} {\bibinfo {author} {\bibfnamefont {B.-L.}\ \bibnamefont
  {Zhang}}, \bibinfo {author} {\bibfnamefont {Z.-H.}\ \bibnamefont {Lei}}, \
  and\ \bibinfo {author} {\bibfnamefont {J.}~\bibnamefont {Tang}},\ }\href@noop
  {} {\  (\bibinfo {year} {2020})},\ \Eprint {http://arxiv.org/abs/2008.07116}
  {arXiv:2008.07116 [hep-ph]} \BibitemShut {NoStop}%
\bibitem [{\citenamefont {Boschini}\ \emph {et~al.}(2018)\citenamefont
  {Boschini} \emph {et~al.}}]{Boschini:2018zdv}%
  \BibitemOpen
  \bibfield  {author} {\bibinfo {author} {\bibfnamefont {M.}~\bibnamefont
  {Boschini}} \emph {et~al.},\ }\href {\doibase 10.3847/1538-4357/aaa75e}
  {\bibfield  {journal} {\bibinfo  {journal} {Astrophys. J.}\ }\textbf
  {\bibinfo {volume} {854}},\ \bibinfo {pages} {94} (\bibinfo {year} {2018})},\
  \Eprint {http://arxiv.org/abs/1801.04059} {arXiv:1801.04059 [astro-ph.HE]}
  \BibitemShut {NoStop}%
\bibitem [{\citenamefont {Bisschoff}\ \emph {et~al.}(2019)\citenamefont
  {Bisschoff}, \citenamefont {Potgieter},\ and\ \citenamefont
  {Aslam}}]{Bisschoff:2019lne}%
  \BibitemOpen
  \bibfield  {author} {\bibinfo {author} {\bibfnamefont {D.}~\bibnamefont
  {Bisschoff}}, \bibinfo {author} {\bibfnamefont {M.~S.}\ \bibnamefont
  {Potgieter}}, \ and\ \bibinfo {author} {\bibfnamefont {O.~P.~M.}\
  \bibnamefont {Aslam}},\ }\href@noop {} {\  (\bibinfo {year} {2019})},\
  \Eprint {http://arxiv.org/abs/1902.10438} {arXiv:1902.10438 [astro-ph.HE]}
  \BibitemShut {NoStop}%
\bibitem [{\citenamefont {Cummings}\ \emph {et~al.}(2016)\citenamefont
  {Cummings}, \citenamefont {Stone}, \citenamefont {Heikkila}, \citenamefont
  {Lal}, \citenamefont {Webber}, \citenamefont {J{\'{o}}hannesson},
  \citenamefont {Moskalenko}, \citenamefont {Orlando},\ and\ \citenamefont
  {Porter}}]{Cummings_2016}%
  \BibitemOpen
  \bibfield  {author} {\bibinfo {author} {\bibfnamefont {A.~C.}\ \bibnamefont
  {Cummings}}, \bibinfo {author} {\bibfnamefont {E.~C.}\ \bibnamefont {Stone}},
  \bibinfo {author} {\bibfnamefont {B.~C.}\ \bibnamefont {Heikkila}}, \bibinfo
  {author} {\bibfnamefont {N.}~\bibnamefont {Lal}}, \bibinfo {author}
  {\bibfnamefont {W.~R.}\ \bibnamefont {Webber}}, \bibinfo {author}
  {\bibfnamefont {G.}~\bibnamefont {J{\'{o}}hannesson}}, \bibinfo {author}
  {\bibfnamefont {I.~V.}\ \bibnamefont {Moskalenko}}, \bibinfo {author}
  {\bibfnamefont {E.}~\bibnamefont {Orlando}}, \ and\ \bibinfo {author}
  {\bibfnamefont {T.~A.}\ \bibnamefont {Porter}},\ }\href {\doibase
  10.3847/0004-637x/831/1/18} {\bibfield  {journal} {\bibinfo  {journal} {The
  Astrophysical Journal}\ }\textbf {\bibinfo {volume} {831}},\ \bibinfo {pages}
  {18} (\bibinfo {year} {2016})}\BibitemShut {NoStop}%
\bibitem [{\citenamefont {Aguilar}\ \emph {et~al.}(2014)\citenamefont {Aguilar}
  \emph {et~al.}}]{Aguilar:2014fea}%
  \BibitemOpen
  \bibfield  {author} {\bibinfo {author} {\bibfnamefont {M.}~\bibnamefont
  {Aguilar}} \emph {et~al.} (\bibinfo {collaboration} {AMS}),\ }\href {\doibase
  10.1103/PhysRevLett.113.221102} {\bibfield  {journal} {\bibinfo  {journal}
  {Phys. Rev. Lett.}\ }\textbf {\bibinfo {volume} {113}},\ \bibinfo {pages}
  {221102} (\bibinfo {year} {2014})}\BibitemShut {NoStop}%
\bibitem [{\citenamefont {Adriani}\ \emph {et~al.}(2011)\citenamefont {Adriani}
  \emph {et~al.}}]{Adriani:2011xv}%
  \BibitemOpen
  \bibfield  {author} {\bibinfo {author} {\bibfnamefont {O.}~\bibnamefont
  {Adriani}} \emph {et~al.} (\bibinfo {collaboration} {PAMELA}),\ }\href
  {\doibase 10.1103/PhysRevLett.106.201101} {\bibfield  {journal} {\bibinfo
  {journal} {Phys. Rev. Lett.}\ }\textbf {\bibinfo {volume} {106}},\ \bibinfo
  {pages} {201101} (\bibinfo {year} {2011})},\ \Eprint
  {http://arxiv.org/abs/1103.2880} {arXiv:1103.2880 [astro-ph.HE]} \BibitemShut
  {NoStop}%
\bibitem [{\citenamefont {Bays}\ \emph {et~al.}(2012)\citenamefont {Bays} \emph
  {et~al.}}]{Bays:2011si}%
  \BibitemOpen
  \bibfield  {author} {\bibinfo {author} {\bibfnamefont {K.}~\bibnamefont
  {Bays}} \emph {et~al.} (\bibinfo {collaboration} {Super-Kamiokande}),\ }\href
  {\doibase 10.1103/PhysRevD.85.052007} {\bibfield  {journal} {\bibinfo
  {journal} {Phys. Rev. D}\ }\textbf {\bibinfo {volume} {85}},\ \bibinfo
  {pages} {052007} (\bibinfo {year} {2012})},\ \Eprint
  {http://arxiv.org/abs/1111.5031} {arXiv:1111.5031 [hep-ex]} \BibitemShut
  {NoStop}%
\bibitem [{\citenamefont {Essig}\ \emph {et~al.}(2012)\citenamefont {Essig},
  \citenamefont {Mardon},\ and\ \citenamefont {Volansky}}]{Essig:2011nj}%
  \BibitemOpen
  \bibfield  {author} {\bibinfo {author} {\bibfnamefont {R.}~\bibnamefont
  {Essig}}, \bibinfo {author} {\bibfnamefont {J.}~\bibnamefont {Mardon}}, \
  and\ \bibinfo {author} {\bibfnamefont {T.}~\bibnamefont {Volansky}},\ }\href
  {\doibase 10.1103/PhysRevD.85.076007} {\bibfield  {journal} {\bibinfo
  {journal} {Phys. Rev. D}\ }\textbf {\bibinfo {volume} {85}},\ \bibinfo
  {pages} {076007} (\bibinfo {year} {2012})},\ \Eprint
  {http://arxiv.org/abs/1108.5383} {arXiv:1108.5383 [hep-ph]} \BibitemShut
  {NoStop}%
\bibitem [{\citenamefont {Essig}\ \emph {et~al.}(2016)\citenamefont {Essig},
  \citenamefont {Fernandez-Serra}, \citenamefont {Mardon}, \citenamefont
  {Soto}, \citenamefont {Volansky},\ and\ \citenamefont {Yu}}]{Essig:2015cda}%
  \BibitemOpen
  \bibfield  {author} {\bibinfo {author} {\bibfnamefont {R.}~\bibnamefont
  {Essig}}, \bibinfo {author} {\bibfnamefont {M.}~\bibnamefont
  {Fernandez-Serra}}, \bibinfo {author} {\bibfnamefont {J.}~\bibnamefont
  {Mardon}}, \bibinfo {author} {\bibfnamefont {A.}~\bibnamefont {Soto}},
  \bibinfo {author} {\bibfnamefont {T.}~\bibnamefont {Volansky}}, \ and\
  \bibinfo {author} {\bibfnamefont {T.-T.}\ \bibnamefont {Yu}},\ }\href
  {\doibase 10.1007/JHEP05(2016)046} {\bibfield  {journal} {\bibinfo  {journal}
  {JHEP}\ }\textbf {\bibinfo {volume} {05}},\ \bibinfo {pages} {046} (\bibinfo
  {year} {2016})},\ \Eprint {http://arxiv.org/abs/1509.01598} {arXiv:1509.01598
  [hep-ph]} \BibitemShut {NoStop}%
\bibitem [{\citenamefont {Trickle}\ \emph {et~al.}(2020)\citenamefont
  {Trickle}, \citenamefont {Zhang},\ and\ \citenamefont
  {Zurek}}]{Trickle:2019ovy}%
  \BibitemOpen
  \bibfield  {author} {\bibinfo {author} {\bibfnamefont {T.}~\bibnamefont
  {Trickle}}, \bibinfo {author} {\bibfnamefont {Z.}~\bibnamefont {Zhang}}, \
  and\ \bibinfo {author} {\bibfnamefont {K.~M.}\ \bibnamefont {Zurek}},\ }\href
  {\doibase 10.1103/PhysRevLett.124.201801} {\bibfield  {journal} {\bibinfo
  {journal} {Phys. Rev. Lett.}\ }\textbf {\bibinfo {volume} {124}},\ \bibinfo
  {pages} {201801} (\bibinfo {year} {2020})},\ \Eprint
  {http://arxiv.org/abs/1905.13744} {arXiv:1905.13744 [hep-ph]} \BibitemShut
  {NoStop}%
\bibitem [{\citenamefont {Baxter}\ \emph {et~al.}(2020)\citenamefont {Baxter},
  \citenamefont {Kahn},\ and\ \citenamefont {Krnjaic}}]{Baxter:2019pnz}%
  \BibitemOpen
  \bibfield  {author} {\bibinfo {author} {\bibfnamefont {D.}~\bibnamefont
  {Baxter}}, \bibinfo {author} {\bibfnamefont {Y.}~\bibnamefont {Kahn}}, \ and\
  \bibinfo {author} {\bibfnamefont {G.}~\bibnamefont {Krnjaic}},\ }\href
  {\doibase 10.1103/PhysRevD.101.076014} {\bibfield  {journal} {\bibinfo
  {journal} {Phys. Rev. D}\ }\textbf {\bibinfo {volume} {101}},\ \bibinfo
  {pages} {076014} (\bibinfo {year} {2020})},\ \Eprint
  {http://arxiv.org/abs/1908.00012} {arXiv:1908.00012 [hep-ph]} \BibitemShut
  {NoStop}%
\bibitem [{\citenamefont {Liu}\ \emph {et~al.}(2020)\citenamefont {Liu},
  \citenamefont {Wu}, \citenamefont {Chi},\ and\ \citenamefont
  {Chen}}]{Liu:2020pat}%
  \BibitemOpen
  \bibfield  {author} {\bibinfo {author} {\bibfnamefont {C.-P.}\ \bibnamefont
  {Liu}}, \bibinfo {author} {\bibfnamefont {C.-P.}\ \bibnamefont {Wu}},
  \bibinfo {author} {\bibfnamefont {H.-C.}\ \bibnamefont {Chi}}, \ and\
  \bibinfo {author} {\bibfnamefont {J.-W.}\ \bibnamefont {Chen}},\ }\href@noop
  {} {\  (\bibinfo {year} {2020})},\ \Eprint {http://arxiv.org/abs/2007.10965}
  {arXiv:2007.10965 [hep-ph]} \BibitemShut {NoStop}%
\bibitem [{\citenamefont {Knapen}\ \emph {et~al.}(2017)\citenamefont {Knapen},
  \citenamefont {Lin},\ and\ \citenamefont {Zurek}}]{Knapen:2017xzo}%
  \BibitemOpen
  \bibfield  {author} {\bibinfo {author} {\bibfnamefont {S.}~\bibnamefont
  {Knapen}}, \bibinfo {author} {\bibfnamefont {T.}~\bibnamefont {Lin}}, \ and\
  \bibinfo {author} {\bibfnamefont {K.~M.}\ \bibnamefont {Zurek}},\ }\href
  {\doibase 10.1103/PhysRevD.96.115021} {\bibfield  {journal} {\bibinfo
  {journal} {Phys. Rev. D}\ }\textbf {\bibinfo {volume} {96}},\ \bibinfo
  {pages} {115021} (\bibinfo {year} {2017})},\ \Eprint
  {http://arxiv.org/abs/1709.07882} {arXiv:1709.07882 [hep-ph]} \BibitemShut
  {NoStop}%
\bibitem [{\citenamefont {Aprile}\ \emph {et~al.}(2019)\citenamefont {Aprile}
  \emph {et~al.}}]{Aprile:2019xxb}%
  \BibitemOpen
  \bibfield  {author} {\bibinfo {author} {\bibfnamefont {E.}~\bibnamefont
  {Aprile}} \emph {et~al.} (\bibinfo {collaboration} {XENON}),\ }\href
  {\doibase 10.1103/PhysRevLett.123.251801} {\bibfield  {journal} {\bibinfo
  {journal} {Phys. Rev. Lett.}\ }\textbf {\bibinfo {volume} {123}},\ \bibinfo
  {pages} {251801} (\bibinfo {year} {2019})},\ \Eprint
  {http://arxiv.org/abs/1907.11485} {arXiv:1907.11485 [hep-ex]} \BibitemShut
  {NoStop}%
\bibitem [{\citenamefont {Acciarri}\ \emph {et~al.}(2015)\citenamefont
  {Acciarri} \emph {et~al.}}]{Acciarri:2015uup}%
  \BibitemOpen
  \bibfield  {author} {\bibinfo {author} {\bibfnamefont {R.}~\bibnamefont
  {Acciarri}} \emph {et~al.} (\bibinfo {collaboration} {DUNE}),\ }\href@noop {}
  {\  (\bibinfo {year} {2015})},\ \Eprint {http://arxiv.org/abs/1512.06148}
  {arXiv:1512.06148 [physics.ins-det]} \BibitemShut {NoStop}%
\bibitem [{\citenamefont {Necib}\ \emph {et~al.}(2017)\citenamefont {Necib},
  \citenamefont {Moon}, \citenamefont {Wongjirad},\ and\ \citenamefont
  {Conrad}}]{Necib:2016aez}%
  \BibitemOpen
  \bibfield  {author} {\bibinfo {author} {\bibfnamefont {L.}~\bibnamefont
  {Necib}}, \bibinfo {author} {\bibfnamefont {J.}~\bibnamefont {Moon}},
  \bibinfo {author} {\bibfnamefont {T.}~\bibnamefont {Wongjirad}}, \ and\
  \bibinfo {author} {\bibfnamefont {J.~M.}\ \bibnamefont {Conrad}},\ }\href
  {\doibase 10.1103/PhysRevD.95.075018} {\bibfield  {journal} {\bibinfo
  {journal} {Phys. Rev. D}\ }\textbf {\bibinfo {volume} {95}},\ \bibinfo
  {pages} {075018} (\bibinfo {year} {2017})},\ \Eprint
  {http://arxiv.org/abs/1610.03486} {arXiv:1610.03486 [hep-ph]} \BibitemShut
  {NoStop}%
\bibitem [{\citenamefont {Khoury}\ and\ \citenamefont
  {Weltman}(2004)}]{Khoury:2003aq}%
  \BibitemOpen
  \bibfield  {author} {\bibinfo {author} {\bibfnamefont {J.}~\bibnamefont
  {Khoury}}\ and\ \bibinfo {author} {\bibfnamefont {A.}~\bibnamefont
  {Weltman}},\ }\href {\doibase 10.1103/PhysRevLett.93.171104} {\bibfield
  {journal} {\bibinfo  {journal} {Phys. Rev. Lett.}\ }\textbf {\bibinfo
  {volume} {93}},\ \bibinfo {pages} {171104} (\bibinfo {year} {2004})},\
  \Eprint {http://arxiv.org/abs/astro-ph/0309300} {arXiv:astro-ph/0309300
  [astro-ph]} \BibitemShut {NoStop}%
\bibitem [{\citenamefont {Nelson}\ and\ \citenamefont
  {Walsh}(2008{\natexlab{a}})}]{Nelson_2008}%
  \BibitemOpen
  \bibfield  {author} {\bibinfo {author} {\bibfnamefont {A.~E.}\ \bibnamefont
  {Nelson}}\ and\ \bibinfo {author} {\bibfnamefont {J.}~\bibnamefont {Walsh}},\
  }\href {\doibase 10.1103/physrevd.77.033001} {\bibfield  {journal} {\bibinfo
  {journal} {Physical Review D}\ }\textbf {\bibinfo {volume} {77}} (\bibinfo
  {year} {2008}{\natexlab{a}}),\ 10.1103/physrevd.77.033001}\BibitemShut
  {NoStop}%
\bibitem [{\citenamefont {Nelson}\ and\ \citenamefont
  {Walsh}(2008{\natexlab{b}})}]{Nelson:2008tn}%
  \BibitemOpen
  \bibfield  {author} {\bibinfo {author} {\bibfnamefont {A.}~\bibnamefont
  {Nelson}}\ and\ \bibinfo {author} {\bibfnamefont {J.}~\bibnamefont {Walsh}},\
  }\href {\doibase 10.1103/PhysRevD.77.095006} {\bibfield  {journal} {\bibinfo
  {journal} {Phys. Rev. D}\ }\textbf {\bibinfo {volume} {77}},\ \bibinfo
  {pages} {095006} (\bibinfo {year} {2008}{\natexlab{b}})},\ \Eprint
  {http://arxiv.org/abs/0802.0762} {arXiv:0802.0762 [hep-ph]} \BibitemShut
  {NoStop}%
\bibitem [{\citenamefont {DeRocco}\ \emph {et~al.}(2020)\citenamefont
  {DeRocco}, \citenamefont {Graham},\ and\ \citenamefont
  {Rajendran}}]{DeRocco:2020xdt}%
  \BibitemOpen
  \bibfield  {author} {\bibinfo {author} {\bibfnamefont {W.}~\bibnamefont
  {DeRocco}}, \bibinfo {author} {\bibfnamefont {P.~W.}\ \bibnamefont {Graham}},
  \ and\ \bibinfo {author} {\bibfnamefont {S.}~\bibnamefont {Rajendran}},\
  }\href@noop {} {\  (\bibinfo {year} {2020})},\ \Eprint
  {http://arxiv.org/abs/2006.15112} {arXiv:2006.15112 [hep-ph]} \BibitemShut
  {NoStop}%
\bibitem [{\citenamefont {Bordag}\ \emph {et~al.}(2001)\citenamefont {Bordag},
  \citenamefont {Mohideen},\ and\ \citenamefont
  {Mostepanenko}}]{Bordag:2001qi}%
  \BibitemOpen
  \bibfield  {author} {\bibinfo {author} {\bibfnamefont {M.}~\bibnamefont
  {Bordag}}, \bibinfo {author} {\bibfnamefont {U.}~\bibnamefont {Mohideen}}, \
  and\ \bibinfo {author} {\bibfnamefont {V.}~\bibnamefont {Mostepanenko}},\
  }\href {\doibase 10.1016/S0370-1573(01)00015-1} {\bibfield  {journal}
  {\bibinfo  {journal} {Phys. Rept.}\ }\textbf {\bibinfo {volume} {353}},\
  \bibinfo {pages} {1} (\bibinfo {year} {2001})},\ \Eprint
  {http://arxiv.org/abs/quant-ph/0106045} {arXiv:quant-ph/0106045} \BibitemShut
  {NoStop}%
\bibitem [{\citenamefont {Adelberger}\ \emph {et~al.}(2007)\citenamefont
  {Adelberger}, \citenamefont {Heckel}, \citenamefont {Hoedl}, \citenamefont
  {Hoyle}, \citenamefont {Kapner},\ and\ \citenamefont
  {Upadhye}}]{Adelberger:2006dh}%
  \BibitemOpen
  \bibfield  {author} {\bibinfo {author} {\bibfnamefont {E.}~\bibnamefont
  {Adelberger}}, \bibinfo {author} {\bibfnamefont {B.~R.}\ \bibnamefont
  {Heckel}}, \bibinfo {author} {\bibfnamefont {S.~A.}\ \bibnamefont {Hoedl}},
  \bibinfo {author} {\bibfnamefont {C.}~\bibnamefont {Hoyle}}, \bibinfo
  {author} {\bibfnamefont {D.}~\bibnamefont {Kapner}}, \ and\ \bibinfo {author}
  {\bibfnamefont {A.}~\bibnamefont {Upadhye}},\ }\href {\doibase
  10.1103/PhysRevLett.98.131104} {\bibfield  {journal} {\bibinfo  {journal}
  {Phys. Rev. Lett.}\ }\textbf {\bibinfo {volume} {98}},\ \bibinfo {pages}
  {131104} (\bibinfo {year} {2007})},\ \Eprint
  {http://arxiv.org/abs/hep-ph/0611223} {arXiv:hep-ph/0611223} \BibitemShut
  {NoStop}%
\bibitem [{\citenamefont {Adelberger}\ \emph {et~al.}(2009)\citenamefont
  {Adelberger}, \citenamefont {Gundlach}, \citenamefont {Heckel}, \citenamefont
  {Hoedl},\ and\ \citenamefont {Schlamminger}}]{Adelberger:2009zz}%
  \BibitemOpen
  \bibfield  {author} {\bibinfo {author} {\bibfnamefont {E.}~\bibnamefont
  {Adelberger}}, \bibinfo {author} {\bibfnamefont {J.}~\bibnamefont
  {Gundlach}}, \bibinfo {author} {\bibfnamefont {B.}~\bibnamefont {Heckel}},
  \bibinfo {author} {\bibfnamefont {S.}~\bibnamefont {Hoedl}}, \ and\ \bibinfo
  {author} {\bibfnamefont {S.}~\bibnamefont {Schlamminger}},\ }\href {\doibase
  10.1016/j.ppnp.2008.08.002} {\bibfield  {journal} {\bibinfo  {journal} {Prog.
  Part. Nucl. Phys.}\ }\textbf {\bibinfo {volume} {62}},\ \bibinfo {pages}
  {102} (\bibinfo {year} {2009})}\BibitemShut {NoStop}%
\bibitem [{\citenamefont {Bjorken}\ \emph {et~al.}(1988)\citenamefont
  {Bjorken}, \citenamefont {Ecklund}, \citenamefont {Nelson}, \citenamefont
  {Abashian}, \citenamefont {Church}, \citenamefont {Lu}, \citenamefont {Mo},
  \citenamefont {Nunamaker},\ and\ \citenamefont {Rassmann}}]{Bjorken:1988as}%
  \BibitemOpen
  \bibfield  {author} {\bibinfo {author} {\bibfnamefont {J.}~\bibnamefont
  {Bjorken}}, \bibinfo {author} {\bibfnamefont {S.}~\bibnamefont {Ecklund}},
  \bibinfo {author} {\bibfnamefont {W.}~\bibnamefont {Nelson}}, \bibinfo
  {author} {\bibfnamefont {A.}~\bibnamefont {Abashian}}, \bibinfo {author}
  {\bibfnamefont {C.}~\bibnamefont {Church}}, \bibinfo {author} {\bibfnamefont
  {B.}~\bibnamefont {Lu}}, \bibinfo {author} {\bibfnamefont {L.}~\bibnamefont
  {Mo}}, \bibinfo {author} {\bibfnamefont {T.}~\bibnamefont {Nunamaker}}, \
  and\ \bibinfo {author} {\bibfnamefont {P.}~\bibnamefont {Rassmann}},\ }\href
  {\doibase 10.1103/PhysRevD.38.3375} {\bibfield  {journal} {\bibinfo
  {journal} {Phys. Rev. D}\ }\textbf {\bibinfo {volume} {38}},\ \bibinfo
  {pages} {3375} (\bibinfo {year} {1988})}\BibitemShut {NoStop}%
\bibitem [{\citenamefont {Davier}\ and\ \citenamefont
  {Nguyen~Ngoc}(1989)}]{Davier:1989wz}%
  \BibitemOpen
  \bibfield  {author} {\bibinfo {author} {\bibfnamefont {M.}~\bibnamefont
  {Davier}}\ and\ \bibinfo {author} {\bibfnamefont {H.}~\bibnamefont
  {Nguyen~Ngoc}},\ }\href {\doibase 10.1016/0370-2693(89)90174-3} {\bibfield
  {journal} {\bibinfo  {journal} {Phys. Lett. B}\ }\textbf {\bibinfo {volume}
  {229}},\ \bibinfo {pages} {150} (\bibinfo {year} {1989})}\BibitemShut
  {NoStop}%
\bibitem [{\citenamefont {Andreas}\ \emph {et~al.}(2012)\citenamefont
  {Andreas}, \citenamefont {Niebuhr},\ and\ \citenamefont
  {Ringwald}}]{Andreas:2012mt}%
  \BibitemOpen
  \bibfield  {author} {\bibinfo {author} {\bibfnamefont {S.}~\bibnamefont
  {Andreas}}, \bibinfo {author} {\bibfnamefont {C.}~\bibnamefont {Niebuhr}}, \
  and\ \bibinfo {author} {\bibfnamefont {A.}~\bibnamefont {Ringwald}},\ }\href
  {\doibase 10.1103/PhysRevD.86.095019} {\bibfield  {journal} {\bibinfo
  {journal} {Phys. Rev. D}\ }\textbf {\bibinfo {volume} {86}},\ \bibinfo
  {pages} {095019} (\bibinfo {year} {2012})},\ \Eprint
  {http://arxiv.org/abs/1209.6083} {arXiv:1209.6083 [hep-ph]} \BibitemShut
  {NoStop}%
\bibitem [{\citenamefont {Banerjee}\ \emph {et~al.}(2019)\citenamefont
  {Banerjee} \emph {et~al.}}]{NA64:2019imj}%
  \BibitemOpen
  \bibfield  {author} {\bibinfo {author} {\bibfnamefont {D.}~\bibnamefont
  {Banerjee}} \emph {et~al.},\ }\href {\doibase 10.1103/PhysRevLett.123.121801}
  {\bibfield  {journal} {\bibinfo  {journal} {Phys. Rev. Lett.}\ }\textbf
  {\bibinfo {volume} {123}},\ \bibinfo {pages} {121801} (\bibinfo {year}
  {2019})},\ \Eprint {http://arxiv.org/abs/1906.00176} {arXiv:1906.00176
  [hep-ex]} \BibitemShut {NoStop}%
\bibitem [{\citenamefont {Lees}\ \emph {et~al.}(2017)\citenamefont {Lees} \emph
  {et~al.}}]{Lees:2017lec}%
  \BibitemOpen
  \bibfield  {author} {\bibinfo {author} {\bibfnamefont {J.}~\bibnamefont
  {Lees}} \emph {et~al.} (\bibinfo {collaboration} {BaBar}),\ }\href {\doibase
  10.1103/PhysRevLett.119.131804} {\bibfield  {journal} {\bibinfo  {journal}
  {Phys. Rev. Lett.}\ }\textbf {\bibinfo {volume} {119}},\ \bibinfo {pages}
  {131804} (\bibinfo {year} {2017})},\ \Eprint
  {http://arxiv.org/abs/1702.03327} {arXiv:1702.03327 [hep-ex]} \BibitemShut
  {NoStop}%
\bibitem [{\citenamefont {Kahn}\ \emph {et~al.}(2015)\citenamefont {Kahn},
  \citenamefont {Krnjaic}, \citenamefont {Thaler},\ and\ \citenamefont
  {Toups}}]{Kahn:2014sra}%
  \BibitemOpen
  \bibfield  {author} {\bibinfo {author} {\bibfnamefont {Y.}~\bibnamefont
  {Kahn}}, \bibinfo {author} {\bibfnamefont {G.}~\bibnamefont {Krnjaic}},
  \bibinfo {author} {\bibfnamefont {J.}~\bibnamefont {Thaler}}, \ and\ \bibinfo
  {author} {\bibfnamefont {M.}~\bibnamefont {Toups}},\ }\href {\doibase
  10.1103/PhysRevD.91.055006} {\bibfield  {journal} {\bibinfo  {journal} {Phys.
  Rev. D}\ }\textbf {\bibinfo {volume} {91}},\ \bibinfo {pages} {055006}
  (\bibinfo {year} {2015})},\ \Eprint {http://arxiv.org/abs/1411.1055}
  {arXiv:1411.1055 [hep-ph]} \BibitemShut {NoStop}%
\bibitem [{\citenamefont {deNiverville}\ \emph {et~al.}(2011)\citenamefont
  {deNiverville}, \citenamefont {Pospelov},\ and\ \citenamefont
  {Ritz}}]{deNiverville:2011it}%
  \BibitemOpen
  \bibfield  {author} {\bibinfo {author} {\bibfnamefont {P.}~\bibnamefont
  {deNiverville}}, \bibinfo {author} {\bibfnamefont {M.}~\bibnamefont
  {Pospelov}}, \ and\ \bibinfo {author} {\bibfnamefont {A.}~\bibnamefont
  {Ritz}},\ }\href {\doibase 10.1103/PhysRevD.84.075020} {\bibfield  {journal}
  {\bibinfo  {journal} {Phys. Rev. D}\ }\textbf {\bibinfo {volume} {84}},\
  \bibinfo {pages} {075020} (\bibinfo {year} {2011})},\ \Eprint
  {http://arxiv.org/abs/1107.4580} {arXiv:1107.4580 [hep-ph]} \BibitemShut
  {NoStop}%
\bibitem [{\citenamefont {Batell}\ \emph {et~al.}(2009)\citenamefont {Batell},
  \citenamefont {Pospelov},\ and\ \citenamefont {Ritz}}]{Batell:2009di}%
  \BibitemOpen
  \bibfield  {author} {\bibinfo {author} {\bibfnamefont {B.}~\bibnamefont
  {Batell}}, \bibinfo {author} {\bibfnamefont {M.}~\bibnamefont {Pospelov}}, \
  and\ \bibinfo {author} {\bibfnamefont {A.}~\bibnamefont {Ritz}},\ }\href
  {\doibase 10.1103/PhysRevD.80.095024} {\bibfield  {journal} {\bibinfo
  {journal} {Phys. Rev. D}\ }\textbf {\bibinfo {volume} {80}},\ \bibinfo
  {pages} {095024} (\bibinfo {year} {2009})},\ \Eprint
  {http://arxiv.org/abs/0906.5614} {arXiv:0906.5614 [hep-ph]} \BibitemShut
  {NoStop}%
\bibitem [{\citenamefont {Digman}\ \emph {et~al.}(2019)\citenamefont {Digman},
  \citenamefont {Cappiello}, \citenamefont {Beacom}, \citenamefont {Hirata},\
  and\ \citenamefont {Peter}}]{Digman:2019wdm}%
  \BibitemOpen
  \bibfield  {author} {\bibinfo {author} {\bibfnamefont {M.~C.}\ \bibnamefont
  {Digman}}, \bibinfo {author} {\bibfnamefont {C.~V.}\ \bibnamefont
  {Cappiello}}, \bibinfo {author} {\bibfnamefont {J.~F.}\ \bibnamefont
  {Beacom}}, \bibinfo {author} {\bibfnamefont {C.~M.}\ \bibnamefont {Hirata}},
  \ and\ \bibinfo {author} {\bibfnamefont {A.~H.}\ \bibnamefont {Peter}},\
  }\href {\doibase 10.1103/PhysRevD.100.063013} {\bibfield  {journal} {\bibinfo
   {journal} {Phys. Rev. D}\ }\textbf {\bibinfo {volume} {100}},\ \bibinfo
  {pages} {063013} (\bibinfo {year} {2019})},\ \Eprint
  {http://arxiv.org/abs/1907.10618} {arXiv:1907.10618 [hep-ph]} \BibitemShut
  {NoStop}%
\bibitem [{\citenamefont {Hardy}\ and\ \citenamefont
  {Lasenby}(2017)}]{Hardy:2016kme}%
  \BibitemOpen
  \bibfield  {author} {\bibinfo {author} {\bibfnamefont {E.}~\bibnamefont
  {Hardy}}\ and\ \bibinfo {author} {\bibfnamefont {R.}~\bibnamefont
  {Lasenby}},\ }\href {\doibase 10.1007/JHEP02(2017)033} {\bibfield  {journal}
  {\bibinfo  {journal} {JHEP}\ }\textbf {\bibinfo {volume} {02}},\ \bibinfo
  {pages} {033} (\bibinfo {year} {2017})},\ \Eprint
  {http://arxiv.org/abs/1611.05852} {arXiv:1611.05852 [hep-ph]} \BibitemShut
  {NoStop}%
\bibitem [{\citenamefont {Liu}\ \emph {et~al.}(2016)\citenamefont {Liu},
  \citenamefont {McKeen},\ and\ \citenamefont {Miller}}]{Liu:2016qwd}%
  \BibitemOpen
  \bibfield  {author} {\bibinfo {author} {\bibfnamefont {Y.-S.}\ \bibnamefont
  {Liu}}, \bibinfo {author} {\bibfnamefont {D.}~\bibnamefont {McKeen}}, \ and\
  \bibinfo {author} {\bibfnamefont {G.~A.}\ \bibnamefont {Miller}},\ }\href
  {\doibase 10.1103/PhysRevLett.117.101801} {\bibfield  {journal} {\bibinfo
  {journal} {Phys. Rev. Lett.}\ }\textbf {\bibinfo {volume} {117}},\ \bibinfo
  {pages} {101801} (\bibinfo {year} {2016})},\ \Eprint
  {http://arxiv.org/abs/1605.04612} {arXiv:1605.04612 [hep-ph]} \BibitemShut
  {NoStop}%
\bibitem [{\citenamefont {Barak}\ \emph {et~al.}(2020)\citenamefont {Barak}
  \emph {et~al.}}]{Barak:2020fql}%
  \BibitemOpen
  \bibfield  {author} {\bibinfo {author} {\bibfnamefont {L.}~\bibnamefont
  {Barak}} \emph {et~al.} (\bibinfo {collaboration} {SENSEI}),\ }\href@noop {}
  {\  (\bibinfo {year} {2020})},\ \Eprint {http://arxiv.org/abs/2004.11378}
  {arXiv:2004.11378 [astro-ph.CO]} \BibitemShut {NoStop}%
\bibitem [{\citenamefont {Krnjaic}\ and\ \citenamefont
  {McDermott}(2020)}]{Krnjaic:2019dzc}%
  \BibitemOpen
  \bibfield  {author} {\bibinfo {author} {\bibfnamefont {G.}~\bibnamefont
  {Krnjaic}}\ and\ \bibinfo {author} {\bibfnamefont {S.~D.}\ \bibnamefont
  {McDermott}},\ }\href {\doibase 10.1103/PhysRevD.101.123022} {\bibfield
  {journal} {\bibinfo  {journal} {Phys. Rev. D}\ }\textbf {\bibinfo {volume}
  {101}},\ \bibinfo {pages} {123022} (\bibinfo {year} {2020})},\ \Eprint
  {http://arxiv.org/abs/1908.00007} {arXiv:1908.00007 [hep-ph]} \BibitemShut
  {NoStop}%
\bibitem [{\citenamefont {Sabti}\ \emph {et~al.}(2020)\citenamefont {Sabti},
  \citenamefont {Alvey}, \citenamefont {Escudero}, \citenamefont {Fairbairn},\
  and\ \citenamefont {Blas}}]{Sabti:2019mhn}%
  \BibitemOpen
  \bibfield  {author} {\bibinfo {author} {\bibfnamefont {N.}~\bibnamefont
  {Sabti}}, \bibinfo {author} {\bibfnamefont {J.}~\bibnamefont {Alvey}},
  \bibinfo {author} {\bibfnamefont {M.}~\bibnamefont {Escudero}}, \bibinfo
  {author} {\bibfnamefont {M.}~\bibnamefont {Fairbairn}}, \ and\ \bibinfo
  {author} {\bibfnamefont {D.}~\bibnamefont {Blas}},\ }\href {\doibase
  10.1088/1475-7516/2020/01/004} {\bibfield  {journal} {\bibinfo  {journal}
  {JCAP}\ }\textbf {\bibinfo {volume} {01}},\ \bibinfo {pages} {004} (\bibinfo
  {year} {2020})},\ \Eprint {http://arxiv.org/abs/1910.01649} {arXiv:1910.01649
  [hep-ph]} \BibitemShut {NoStop}%
\bibitem [{\citenamefont {Escudero}(2019)}]{Escudero:2018mvt}%
  \BibitemOpen
  \bibfield  {author} {\bibinfo {author} {\bibfnamefont {M.}~\bibnamefont
  {Escudero}},\ }\href {\doibase 10.1088/1475-7516/2019/02/007} {\bibfield
  {journal} {\bibinfo  {journal} {JCAP}\ }\textbf {\bibinfo {volume} {02}},\
  \bibinfo {pages} {007} (\bibinfo {year} {2019})},\ \Eprint
  {http://arxiv.org/abs/1812.05605} {arXiv:1812.05605 [hep-ph]} \BibitemShut
  {NoStop}%
\end{thebibliography}%

\end{document}